\documentclass[aps,prd,10pt,showpacs,amsmath,showkeys,twocolumn,floatfix,amssymb, preprintnumbers, nofootinbib, superscriptaddress]{revtex4-1} 
\usepackage{epsfig,dcolumn}
\usepackage{graphicx}
\usepackage{comment} 
\usepackage[usenames]{color}
\usepackage{graphicx}
\usepackage{bm}
 \usepackage{ifpdf}

\usepackage[normalem]{ulem}
\usepackage[dvipsnames]{xcolor}
\usepackage[utf8]{inputenc}
\usepackage{hyperref}
\hypersetup{ 
  pdfnewwindow=true,      
  colorlinks=true,        
  linkcolor=PineGreen,    
  citecolor=PineGreen,    
  filecolor=PineGreen,    
  urlcolor=PineGreen      
}

\newcommand{\be}{\begin{eqnarray}}
\newcommand{\ee}{\end{eqnarray}}

\newcommand{\cor}[2]{{\color{red} \sout{#1}} {\color{blue} #2}}

\begin{document}

\title{ Variational approach to $N$-body interactions in  finite volume     }

\author{Peng~Guo}
\email{pguo@csub.edu}
\affiliation{Department of Physics and Engineering,  California State University, Bakersfield, CA 93311, USA.}
\affiliation{Kavli Institute for Theoretical Physics, University of California, Santa Barbara, CA 93106, USA}

\author{Michael~D\"oring}
\affiliation{Department of Physics, The George Washington University, Washington, DC 20052, USA}
\affiliation{Theory Center, Thomas Jefferson National Accelerator Facility, Newport News, VA 23606, USA}

\author{Adam~P.~Szczepaniak}
\affiliation{Physics Department, Indiana University, Bloomington, IN 47405, USA}
\affiliation{Center for Exploration of Energy and Matter, Indiana University, Bloomington, IN 47403, USA}
\affiliation{Theory Center, Thomas Jefferson National Accelerator Facility, Newport News, VA 23606, USA}

\date{\today}

\begin{abstract} 

We explore variational approach to  the finite-volume $N$-body  problem. The general formalism  for N    non-relativistic spinless particles interacting with periodic pair-wise potentials  yields N-body secular equations.  The solutions depend on the infinite-volume N-body wave functions.   Given that the infinite-volume N-body dynamics may be solved by the standard Faddeev  approach, the variational N-body formalism can  provide a convenient  numerical framework for finding discrete energy spectra in  periodic lattice structures. 

\end{abstract} 


\preprint{JLAB-THY-18-2837 }

\maketitle

\section{Introduction}

Three-particle channels are important in spectroscopy of  excited mesons. For example, $G$ parity forbids a two pion decay of the $a_1(1260)$ meson so that the width of this axial vector resonance is determined by its coupling to three pions. The three pion decay channel is allowed for the exotic, $J^{PC}=1^{-+}$, $\pi_1$ meson, while the exotic $h_2$ with $J^{PC}=2^{+-}$ could decay to $b_1[\omega\pi]\pi$ and $\rho\pi$. 
 Because of their possible hybrid nature and relation to 
confinement~\cite{gluex} these mesons are a subject of intense theoretical and experimental investigations~\cite{Alekseev:2009aa}. 
In the baryon sector, the situation is even more complex because two and three-particle systems often mix. Phenomenologically it is observed that the $\pi\pi N$ channels have significant influence and cannot be neglected when determining resonance parameters 
of nucleon excitations.  The number of possible quantum numbers ($\pi\Delta^{(*)}$, $\pi N^*$, $\sigma N$, $\rho N$,\dots) for a given spin and parity $J^P$ is large  so that one has to truncate. Yet, phenomenology can be a guideline to select relevant channels, e.g., by inspecting the involved centrifugal barriers. 
For example, it is reasonable to assume that the Roper resonance $N(1440)1/2^+$ is dominated by the $\sigma N$ channel (all three particles in relative $S$-wave) and maybe the $\pi\Delta$ channel, while the $\rho N$ channel could be less important at smaller energies where only pions from the low-energy tail of the $\rho$ can be on-shell.

Pioneering calculations in lattice QCD of the spectra of light excited mesons~\cite{Dudek:2009qf} and baryons~\cite{Edwards:2011jj, Engel:2013ig} deliver a semi-quantitative picture in which pion masses are large and finite-volume effects are usually neglected.
Energy eigenvalues have been calculated for the $a_1(1260)$~\cite{Lang:2014tia} using $q\bar q$ operators but also a $\rho\pi$ meson-meson operator, though taken at zero momentum, $\pi({\bf 0})\,\rho({\bf 0})$. Recently, the HadronSpectrum collaboration calculated isospin $I=2$ $\pi\rho$ scattering although the $\rho$ meson is stable at the pertinent pion mass~\cite{Woss:2018irj}. 

Few-body systems above threshold represent the next milestone for the ab-initio understanding of the strong interaction through lattice QCD calculations. Therefore, the infinite-volume extrapolation of three-body systems has attracted much interest recently~\cite{ Klos:2018sen, Konig:2017krd,Briceno:2018mlh, Briceno:2017tce, Hansen:2016fzj, Hansen:2015zga, Hansen:2014eka, Romero-Lopez:2018rcb, Hammer:2017kms, Hammer:2017uqm,  Meissner:2014dea,  Polejaeva:2012ut, Mai:2018djl, Doring:2018xxx, Mai:2017bge,  Briceno:2012rv, Kreuzer:2012sr,Guo:2016fgl,Guo:2017ism,Guo:2017crd,Guo:2018xbv,Agadjanov:2016mao,Hansen:2017mnd,Orasch:2018xxz}, including an extension to coupled $2$ to $3$-body channels~\cite{Briceno:2017tce} and an study of connection between low temperature condensation and scattering in lattice $\phi^4$ theory \cite{Orasch:2018xxz}.
For the three-particle system in the finite volume, the energy eigenvalues are expected to change not only quantitatively but also qualitatively above  threshold, compared to the two-body case.
One can easily understand this by considering the non-interacting energies. For two particles of equal mass $m$ in a cube of side lengths $L$ with periodic boundary conditions, they are given by $E=2E_{\bf n}$ where $E_{\bf n}=\sqrt{m^2+{\bf p_{\bf n}}^2}$ and ${\bf p_n}=(2\pi/L)\,{\bf n}$, ${\bf n}\in \mathbb{Z}^3$. 
In contrast, noninteracting three-body energies are given by $E=E_{\bf n}+E_{{\bf n}'}+E_{{\bf n+n}'}$ where ${\bf n,n}'\in \mathbb{Z}^3$. This pattern is entirely different from the two-body case which is expected to reflect in the interacting spectrum, as well. 
Apart from conceptual challenges, there is also the problem of underdetermination, {\it i.e.} there being a plethora of possible channels  in contrast to scarce data, and, for the selection of relevant channels, similar considerations as in the infinite volume will have to be made.

Four-body systems above threshold have so many possible combinations of quantum numbers that one can only hope to address the simplest cases. A prime example is the energy region above the four-pion threshold in $\pi\pi$ scattering. In analyses of experiment, usually the four-pion channel below $1\mbox{ GeV}$ is neglected given that the experimentally measured partial wave amplitudes are almost elastic in this energy region (see  Refs.~\cite{Niecknig:2012sj,
GarciaMartin:2011cn} for a calculation of the inelasticity). Similarly, the 
 existing lattice studies of  $\pi\pi$ partial waves 
   exclude the  data above the four pion intermediate states for the infinite-volume extrapolation (see, e.g., Refs.~\cite{Guo:2018zss, Hu:2017wli, Wilson:2015dqa, Bolton:2015psa}). 
   
   So far, in the three-body calculations only two meson-meson or meson-baryon operators are used, and not yet three of type $\pi\pi\pi$ or $\pi\pi N$. An exception is the $N$-body threshold calculation of positively charged pions by the NPLQCD collaboration~\cite{Beane:2007es, Detmold:2008fn}. Energy eigenvalues for the Roper~\cite{Lang:2016hnn} resonance above the $\pi\pi N$ threshold have been calculated recently including non-local $\pi N$ and $\sigma N$ operators. This calculation was performed close to the physical pion mass such that the extracted energy levels lie above the $\pi N$ but also above the $\pi\pi N$ threshold. An unusual pattern was observed that could originate from the afore-mentioned three-body dynamics in coupled $\pi N$, $\sigma N$, $\dots$ channel, where the $\sigma$ has to be understood as an interacting two-body subsystem and not a stable particle.

In a rough and incomplete classification, there are three major approaches to solving the three-body problem in the finite volume, in the momentum space representation:  An relativistic, all-orders perturbation theory 
pursued by Brice\~no, Hansen, Sharpe~\cite{Briceno:2018mlh, Briceno:2017tce, Hansen:2016fzj, Hansen:2015zga, Hansen:2014eka}, a non-relativistic dimer formalism by Hammer, Pang, Rusetsky et al.~\cite{Romero-Lopez:2018rcb, Hammer:2017kms, Hammer:2017uqm,  Meissner:2014dea,  Polejaeva:2012ut} and a method based on three-body unitarity to identify on-shell configurations and, therefore, power-law finite volume-effects by Döring and Mai~\cite{Mai:2018djl, Doring:2018xxx, Mai:2017bge}. For the latter two approaches, the partial diagonalization of the amplitude according to cubic symmetry was discussed in Ref.~\cite{Doring:2018xxx}.

All these approaches aim at fully mapping out the few-body dynamics but there are also attempts to obtain essential information from these system without the need to take explicitly all degrees of freedom into account~\cite{Agadjanov:2016mao, Hansen:2017mnd}. The first-ever prediction of excited three-body energy eigenvalues of a physical system  ($\pi^+\pi^+\pi^+$), from  two-body scattering information and lattice threshold eigenvalues~\cite{Beane:2007es, Detmold:2008fn} was achieved recently~\cite{Mai:2018djl}.


\subsection{Variational approach}
Most of the approaches to the few-body problem, including the examples discussed above, rely on  momentum-representation of the reaction amplitude or correlation function in the finite volume. There is, however, an alternative approach based on Faddeev equations and two- and three-body wave functions in configuration space \cite{Guo:2012hv,Guo:2013vsa,Guo:2016fgl,Guo:2017ism,Guo:2017crd,Guo:2018xbv}. The  finite volume wave function is related to the infinite volume wave function by a  linear superposition over infinite sets of periodic cubic boxes, the quantization conditions are subsequently obtained from  matching conditions   \cite{Guo:2012hv,Guo:2013vsa,Guo:2016fgl,Guo:2017ism,Guo:2017crd,Guo:2018xbv}. One of the advantages of this approach is that the connection between long-distance correlation over boxes and short-distance interaction within each cubic box is made explicit by construction of finite volume wave function.

In the present work, we   set up   and explore a foundation to a potentially convenient  numerical approach to   the $N$-body interaction in finite volume. The formalism presented in this work is based on  the variational method~\cite{Kohn:1948col,Kohn:1954KKR} combined with  the 
Faddeev approach \cite{Faddeev:1960su,Faddeev:1965,Gloeckle:1983,Faddeev:1993}.   A brief summary of the variational principle and Faddeev approach are presented in Appendix \ref{variational} and  \ref{faddeeveq} as a short reference for a reader who is not familiar with above mentioned methods.

Based on the traditional variational principle, the secular equation may be obtained   by considering: \mbox{$\delta \Lambda =0$} with     \mbox{$\Lambda =  \langle \Phi | E- \hat{H}^{(L)} | \Phi\rangle $},  where  both the $N$-body  Hamiltonian $\hat{H}^{(L)} = \hat T + \sum_{i<j}\hat{V}_{(ij)}^{(L)} $ and the trial wave function $\Phi$ display periodicity in the cubic lattice, $\hat T$ is the kinetic energy operator and  $\hat{V}_{(ij)}^{(L)}$ stands for the periodic pair-wise interaction between the $i$-th and $j$-th particle. Instead of the traditional approach, in this work we  
write the total wave function as a sum of multiple terms \cite{Faddeev:1960su,Faddeev:1965,Gloeckle:1983,Faddeev:1993}: \mbox{$ \Phi  = \sum_{(i<j)=1}^{N} \Phi^{(ij)}$}. Each component of the total wave function,  $\Phi^{(ij)}$, is required to satisfy the equation  \mbox{$(  E- \hat{T} ) | \Phi^{(ij)} \rangle =\hat{V}_{(ij)}^{(L)} | \Phi\rangle $}.  Therefore, a single Schr\"odinger equation, \mbox{$(E- \hat{H}^{(L)} ) | \Phi\rangle =0$},  is turned into $N(N-1)/2$   coupled equations. The advantage of Faddeev approach is to allow one to incorporate the dominant subsystem structures in an adequate way and to use two-body scattering amplitudes as input for the $N$-body dynamical equations.  By splitting the complete $N$-body wave function,  ultimately    $N(N-1)/2$   secular equations may be obtained by considering
\begin{equation}
 \delta \left[ \langle \Phi |   (  E- \hat{T} ) | \Phi^{(ij)} \rangle -  \langle \Phi |  \hat{V}_{(ij)}^{(L)} | \Phi\rangle \right ] =0. \label{introsecular}
\end{equation}
 As will be shown in Section \ref{discussion}, the secular equations obtained from Eq.~(\ref{introsecular})  resemble two-body secular equations. 
 
 In solid state and condensed matter physics, the Linear Combination of Atomic Orbital (LCAO) method for the calculations of the electronic structure of periodic systems \cite{Ziman:1964} provides a elegant way to construct a wave function that satisfies periodic boundary conditions. The trial wave function that describes electrons traveling in a periodic crystal is constructed by linear superposition of all atomic orbital solutions of  an isolated atom centered at each cell of the crystal. In this way, the periodic boundary conditions of the wave function is automatically guaranteed, and the energy spectra are given by   secular equations from the variational principle.
 Similarly   as also suggested in Refs.\cite{Guo:2013vsa,Guo:2016fgl,Guo:2017ism,Guo:2017crd},  the trial finite-volume $N$-body wave function    may be constructed from the $N$-body    infinite-volume wave function,  \mbox{$ \Phi (\{ \mathbf{ x}  \})  = \sum_{ \{ \mathbf{ n}_{\mathbf{ x}} \} \in \mathbb{Z}^{3}} \Psi( \{ \mathbf{ x} + \mathbf{ n}_{\mathbf{ x}} L \} )$}, 
 where $\Psi$ is the solution of the corresponding infinite-volume Schr\"odinger equation, \mbox{$(E- \hat{H} ) | \Psi\rangle =0$}, and the set $\{ \mathbf{ x}  \}$ stands for the complete set of particle positions. 
Hence,  the trial finite-volume $N$-body wave function satisfies periodic boundary conditions by construction. In principle, the infinite-volume wave function may be solved by standard methods, such as Faddeev's approach \cite{Faddeev:1960su, Faddeev:1965, Gloeckle:1983, Faddeev:1993}. 

The immediate gain of the variational approach for finite volume systems is  evident.
(1) The construction of the finite-volume wave function from the infinite-volume wave function presents a clear connection  between short-range $N$-body dynamics within each image of the cubic box and long-range correlations in the entire periodic lattice structure. The   
$N$-body dynamics  within each cubic box is  determined  by $N$-body Faddeev equations. The long-range correlation effect accumulated from all cubic boxes is implemented by the linear superposition of all     wave functions   centered at each image of cubic boxes. (2) The quantization conditions (secular equations) due to the periodic structure of the lattice are imposed by a variational approach, and eventually yield the discrete energy spectra of the system in the finite volume.  (3) The  variational formalism is   mathematically  transparent and may be suitable for the numerical evaluation of the $N$-body finite volume problem. However, it comes at the price of sacrificing the explicit analytical expressions of quantization conditions as present in the Lüscher formula \cite{Luscher:1990ux} for the two-body problem and  at the high cost of computation of the $N$-body phase space integration.

As  we present here the first attempt to calculate the $N$-body interaction in finite volume with the proposed methods, and also for the sake of simplification of the discussion, in this work, we only consider a simple model with $N$ non-relativistic spinless particles interacting through pair-wise short-range potentials. Given the   infinite-volume wave function, $\Psi  $, that may be solved by the Faddeev approach and  is used as input to the finite-volume problem, the quantization conditions for the finite volume $N$-body interaction are  obtained and presented in Section \ref{Nbform}. 
In principle, three-body forces and coupled-channel effects may be included in the formalism as well. However,
 such type of effects are not considered in the present work. In addition, bound states below threshold may be  described  by the analytic continuation of scattering amplitudes. In the present work, the focus is only given on the presentation of the $N$-body problem with pair-wise interactions. The discussion of three-body force, coupled-channel effect, and bound states below threshold will be given in the  future publications.

An effective two-body formalism may be accomplished by integrating out the rest of the degrees of freedoms of the $N$-body system.
The resemblance of the $N$-body quantization condition   to the two-body quantization condition will be discussed in Section \ref{Nbeff}. At last, in the Appendix we also present  some main results   for a  pair-wise short-range $\delta$-function potential. The renormalization issue due to the singular nature of the $\delta$-function interaction in $3D$ is also discussed.

The paper is organized as follows. In Section \ref{Nbform} we present the derivation and main results of the variational approach to the $N$-body interaction  in finite volume. The discussion and summary are given in Section \ref{discussion}.


\section{Variational $N$-body interaction formalism  in finite volume}\label{Nbform}

\subsection{Secular equations of $N$-body interaction in finite volume}
 The  dynamics of $N$ non-relativistic particles interacting via  pair-wise  interactions in the finite volume with periodic boundary conditions is determined by the Schr\"odinger  equation,
 \begin{equation}
 \left [ E+  \sum_{k=1}^{N} \frac{\nabla^{2}_{\mathbf{ x}_{k}} }{2 m}   -  \sum_{(i < j)=1}^{N} V^{(L)}_{(ij)} (\mathbf{ x}_{i } - \mathbf{ x}_{j})    \right ]   \Phi( \{ \mathbf{ x}  \}    )=0. \label{fvschr}
\end{equation}
where we take all masses to be equal to $m$,  
 $\mathbf{ x}_{i}$ denotes  
  particle position and   $\{\mathbf{ x}\} \equiv \{\mathbf{ x}_{1},\mathbf{ x}_{2}, \cdots, \mathbf{ x}_{N} \}$.     The     potential between  the  $i$-th and $j$-th particles is described by   \mbox{$ V^{(L)}_{(ij)}$}. In finite volume,   \mbox{$ V^{(L)}_{(ij)}$}   displays periodicity when the  distance of  the $i$-th and $j$-th particles is larger than the size of the cubic box. 
The    \mbox{$ V^{(L)}_{(ij)}$} may be written as the superposition of all the  potentials centered at each image of the periodic cubic box,
\begin{equation}
V^{(L)}_{(ij)} (\mathbf{ r}  ) = \sum_{\mathbf{ n} \in \mathbb{Z}^{3}} V_{(ij)} (\mathbf{ r} + \mathbf{ n} L) ,
\end{equation}
 where     \mbox{$ V_{(ij)}$} is the   potential between  the  $i$-th and the $j$-th particle in  the same box, and $L$ is the size of  the three-dimensional cube. 
 Because of the periodicity of the finite-volume potential, 
 the finite-volume  $N$-particle wave function, $\Phi  (\{ \mathbf{ x}  \})$, must also satisfy the periodic boundary condition,
 \begin{equation}
 \Phi   (\{ \mathbf{ x} + \mathbf{ n}_{\mathbf{ x}} L \} ) = \Phi  (\{ \mathbf{ x} \} ),  
 \end{equation}
 where $\{ \mathbf{ n }_{\mathbf{ x}} \} = \{ \mathbf{ n }_{\mathbf{ x}_1},\mathbf{ n }_{\mathbf{ x}_2}, \cdots, \mathbf{ n }_{\mathbf{ x}_N} \} $ and $\mathbf{ n }_{\mathbf{ x}_i} \in \mathbb{Z}^{3}$.

Following the Faddeev approach  \cite{Faddeev:1960su,Faddeev:1965,Gloeckle:1983,Faddeev:1993}, that is briefly summarized in Appendix \ref{faddeeveq}, the $N$-body finite-volume wave function may be expressed   as the sum of  $N(N-1)/2$ terms,
\begin{equation}
\Phi  (\{ \mathbf{ x} \} )  = \sum_{(i<j)=1}^{N} \Phi^{(ij)}  (\{ \mathbf{ x} \} ) , \label{fvwavsum}
\end{equation}
 with wave function 
  $\Phi^{(ij)}$ required to satisfy the equation	
  \begin{equation}
 \left ( E - \hat{T}       \right )  \Phi^{(ij)}( \{ \mathbf{ x}  \}  )  =  V^{(L)}_{(ij)} (\mathbf{ x}_{i } - \mathbf{ x}_{j})     \Phi ( \{ \mathbf{ x}  \}  ) , \label{phiij}
\end{equation}
where \mbox{$\hat{T} = -\sum_{k=1}^{N} \frac{\nabla^{2}_{\mathbf{ x}_{k}} }{2 m} $}. 
  Hence,   the Schr\"odinger equation, Eq.~(\ref{fvschr}) is converted to $N(N-1)/2$    
coupled equations. 
The $N$-body finite volume wave function is normally expanded in terms of a set of periodic basis functions, 
 \begin{equation}
  \Phi  (\{ \mathbf{ x} \} ) =  \sum_{[J]} c_{[J]}  \Phi_{[J]}  (\{ \mathbf{ x} \} ), \label{exp}
 \end{equation} 
 where $[J]$ refers to a complete set of quantum numbers for the $N$-particle basis  wave functions, and $c_{[J]}$ stands for the expansion coefficients that may be determined by the variational principle. Similar to Eq.~(\ref{fvwavsum}), the basis function $\Phi_{[J]}$ is also given by the sum of $N(N-1)/2$ terms: \mbox{$\Phi_{[J]}  =\sum_{(i<j)=1}^{N} \Phi^{(ij)}_{[J]}  $}. We remark that the choice of basis functions may be arbitrary, depending on the symmetry of the specific physical system and the convenience of numerical computation. However, the different choices  of basis functions should   lead to consistent results. The reasonable choice of basis functions should preserve the symmetry of physical system, such as periodic boundary conditions in finite volume, and numerical results should   be stable and converge.  
 
In the finite volume, to fulfill the periodic boundary condition, the basis functions, $\Phi_{[J]}$, are constructed from   infinite volume solutions of the Schr\"odinger equation. The complete details are given in following Section \ref{wavandsecular}. In the following we use the two-body problem as a specific example to explain  our choice of  basis functions $\Phi_{[J]}$ in finite volume. We denote the infinite volume wave functions by $\Psi_{[J]}$. 
 In the case of two-particles in the center-of-mass frame, the solutions of the Schr\"odinger equation in the infinite volume may be determined by the partial wave expansion of the free incoming wave: $\Psi^{(0)}_{[J]} (\mathbf{ r}) = (4\pi) i^J Y_{[J]} (\mathbf{ r}) j_{J} (q r)$, where $[J]=JM$ are partial wave quantum numbers, $\mathbf{ r}$ stands for the relative position of  the two particles, and \mbox{$q=\sqrt{m E}$} is the relative momentum of the particles. Thus, the asymptotic solution of two-body scattering in infinite volume is given by
\begin{equation}
\Psi_{[J]} (\mathbf{ r}) \rightarrow (4\pi) i^J Y_{[J]} (\mathbf{ r}) \left [ j_J(qr) + i t_J (q) h_J^+ (qr) \right ],
\end{equation}
where $t_J(q)$ stands for the two-body scattering amplitude.
Hence, $\Psi_{[J]}$ can be  
used as basis functions in the infinite volume to construct the corresponding finite-volume basis functions, $\Phi_{[J]}$ in order 
 to  fulfill the requirement of periodic boundary conditions, 
 \begin{equation} 
 \Phi_{[J]} (\mathbf{ r})  = \sum_{\mathbf{n} \in \mathbb{Z}^{3}  } \Psi_{[J]} (\mathbf{ r} + \mathbf{n}  L).   
 \end{equation} 
 Such finite volume basis function  reflect the periodic lattice  structure. It will be shown 
   in Section \ref{reltoLuscher},  that the two-body basis functions,  $\Phi_{[J]}$, will be related to L\"uscher's zeta function \cite{Luscher:1990ux}. As a general remark, the basis functions should be constructed respecting the cubic symmetry of the problem; yet, we are not explicitly addressing this topic in here. Such basis functions will be given by linear superpositions of infinite-volume basis functions; see, for example, Ref.~\cite{Doring:2018xxx} in which basis functions for ``shells'' are derived as linear combinations of cubic harmonics which by themselves are superpositions of spherical harmonics.  

With the finite volume wave function expanded in terms of the basis function, Eq.~(\ref{exp}), the  variational principle, 
 $\partial \Lambda^{(ij)}/\partial c^*_{[J']} =0$, 
with $\Lambda^{(ij)} $ given by 
\begin{align}
 \Lambda^{(ij)} & =  \sum_{[J], [J']} c^*_{[J']} \langle \Phi_{[J']} | \left [  ( E- \hat{T}) | \Phi^{(ij)}_{[J]} \rangle -    \hat{V}^{(L)}_{(ij)} |\Phi_{[J]}  \rangle \right ]c_{[J]}  .
\end{align}
yields a set of coupled secular equations,  
\begin{equation}
 \sum_{[J]} \left [  \langle \Phi_{[J']} \left  | E- \hat{T}  \right    | \Phi^{(ij)}_{[J]} \rangle   -    \langle \Phi_{[J']} \left   |  \hat{V}^{(L)}_{(ij)}  \right  |  \Phi_{[J]} \rangle   \right ]  c_{[J]}  =0,\label{seceq}
\end{equation} 
and summing the above $N(N-1)/2$  equations leads a secular equation of familiar form (see also Eq.~(\ref{eq:seculareasy})),
   \begin{equation}
 \sum_{[J]} \left [  \langle \Phi_{[J']} \left  | E- \hat{H}^{(L)}    \right  |  \Phi_{[J]} \rangle   \right ]  c_{[J]}  =0  ,\label{secvar}
\end{equation} 
where $\hat{H}^{(L)} = \hat{T}  + \sum_{(i<j)=1}^{N} \hat{V}^{(L)}_{(ij)} $. 
A non-trivial solution of Eq.~(\ref{seceq})  exists,  provided all $N(N-1)/2$    determinant conditions,
   \begin{equation}
 \det  \left [  \langle \Phi_{[J']}  \left  | E-  \hat{T} \right    | \Phi^{(ij)}_{[J]} \rangle -   \langle \Phi_{[J']} \left  |  \hat{V}^{(L)}_{(ij)}  \right   |  \Phi_{[J]} \rangle \right ] =0, \label{detfaddeev}
\end{equation}
are satisfied simultaneously. 
To summarize, given a set of 
 basis functions, $\Phi^{(ij)}_{[J]}$ that satisfy periodic boundary conditions, 
the variational solution for the spectrum of   
the $N$-body finite volume problem is reduced to a solution of the determinant conditions given by Eq.~(\ref{detfaddeev}).

\subsection{Construction of finite volume wave function and reduction of the secular equations}\label{wavandsecular}
To proceed,  a proper choice of basis functions, with 
periodic symmetry, 
$\Phi^{(ij)}_{[J]}$, has to be made. 
There exist numerous examples 
 of periodic basis functions, 
{\it e.g.} devised for  calculations of electronic structure in a periodic lattice  and other condensed matter or solid state systems. 
Specifically, in the LCAO method   \cite{Ziman:1964}, the periodic variational basis functions for describing electronic states tunneling in  a crystal are constructed as a  linear superposition of  all    atomic orbital solutions for an isolated atom located 
at each unit cell of the crystal.  
In application to hadron scattering, instead of using bound state solutions, as done previously in ~\cite{Guo:2013vsa,Guo:2016fgl,Guo:2017ism,Guo:2017crd},  the basis functions may be constructed from  the  linear superposition of infinite-volume scattering wave functions centered at each image of the cubic box,
  \begin{equation}
  \Phi^{(ij)}_{[J]}  (\{ \mathbf{ x} \} ) = \frac{1}{\mathcal{N}} \sum_{ \{ \mathbf{ n}_{\mathbf{ x}} \} \in \mathbb{Z}^{3}}   \Psi^{(ij)}_{[J]}  (\{ \mathbf{ x} + \mathbf{ n}_{\mathbf{ x}}  L \} )  , \label{fvwavijconstr}
 \end{equation} 
 where $\{ \mathbf{ n}_{\mathbf{ x}} \}  = \{ \mathbf{ n}_{\mathbf{ x}_{1}},\mathbf{ n}_{\mathbf{ x}_{2}}, \cdots , \mathbf{ n}_{\mathbf{ x}_{N}} \} $,  \mbox{$\mathcal{N} = \sum_{\mathbf{ n} \in \mathbb{Z}^{3}}$} is a normalization factor, and $ \Psi^{(ij)}_{[J]} $ satisfies the  Schr\"odinger  equation in  the infinite volume with  potentials $V_{(ij)}$,
   \begin{equation}
   ( E -\hat{T}        ) |   \Psi^{(ij)}_{[J]} \rangle =  \hat{V}_{(ij)}  |   \Psi_{[J]}  \rangle . \label{Psiij}
\end{equation}
The  total $N$-body wave function in  the infinite volume,
$\Psi_{[J]}$, is the scattering solution of the Schr\"odinger  equation, \mbox{$(E-\hat{T}-\hat{V} ) |   \Psi_{[J]}  \rangle=0$}, and  may be expressed as  
\begin{equation}
\Psi_{[J]} =\Psi^{(0)}_{[J]}   +\sum_{(i<j)=1}^{N} \Psi^{(ij)}_{[J]},
\end{equation}
where $\Psi^{(0)}_{[J]} $ stands for the free incoming wave and satisfies the free Schr\"odinger equation: $( E -\hat{T}        ) |   \Psi^{(0)}_{[J]} \rangle=0$. For example, the two-body    
 partial wave   free incoming wave are given by   $\Psi^{(0)}_{[J]}( \mathbf{r}) = (4\pi) i^J Y_{[J]} (  \mathbf{r} )j_{J} (q r) $, while the corresponding wave function in the three-body case in given in 
  ~\cite{Guo:2016fgl}. 
   The periodic symmetry of the finite-volume wave function $\Phi$ is satisfied automatically by the construction in Eq.~(\ref{fvwavijconstr}). The  infinite-volume $N$-body wave function, $\Psi$, may be solved by a standard Faddeev approach.  Locally, at each image of the cubic box, assuming the size of the box is large enough,  the short distance  $N$-body dynamics is thus described by the solution of  the infinite-volume Schr\"odinger equation, $\Psi $. The long distance correlations  at scales larger than the range of the potential is taken in to account through the linear superposition of the   infinite-volume wave functions centered at each image of the periodic cubic box, see Eq.~(\ref{fvwavijconstr}).

Using Eqs.~(\ref{fvwavijconstr}),~(\ref{Psiij}) and taking into account periodicity of the finite-volume potential,  the secular equations given by 
Eq.~(\ref{seceq}) reduce to
  \begin{equation}
 \sum_{[J]}  \langle \Phi_{[J']}    |  \hat{V}_{(ij)}    \left [   |    X^{(ij)}_{[J]}  \rangle - | \Phi_{[J]} \rangle   \right ]  c_{[J]}  =0   , \label{rdseceq}
\end{equation} 
where   
  \begin{equation}
 X^{(ij)}_{[J]}   (\{ \mathbf{ x} \} ) = \sum_{ \overline{\{ \mathbf{ n}_{\mathbf{ x}} \}}_{(i,j)} \in \mathbb{Z}^{3}}     \Psi_{[J]}  (\{ \mathbf{ x} + \mathbf{ n}_{\mathbf{ x}}  L \} )  ,\label{xijwavconstr}
 \end{equation} 
 and $ \overline{\{ \mathbf{ n}_{\mathbf{ x}} \}}_{(i,j)}   $ stands for the set $ \{ \mathbf{ n}_{\mathbf{ x}} \} $   excluding two elements: $ \mathbf{ n}_{\mathbf{ x}_{i} }$ and $ \mathbf{ n}_{\mathbf{ x}_{j} }$.
  Furthermore, the $N(N-1)/2$    determinant conditions given by  Eq.(\ref{detfaddeev})  become 
   \begin{equation}
 \det  \left [  \langle \Phi_{[J']}    |  \hat{V}_{(ij)}    \left (     |      X^{(ij)}_{[J]}       \rangle - | \Phi_{[J]} \rangle   \right )  \right ] =0. \label{rddet}
\end{equation}
Given the scattering solution of the $N$-body problem in infinite volume, $\Psi$, as input, the  finite volume wave functions, $\Phi$ and $X^{(ij)}$, can be constructed by using 
Eq.~(\ref{fvwavijconstr}) and Eq.~(\ref{xijwavconstr}) respectively. 
  We remark that  the energy dependence of the infinite volume wave function, $\Psi_{[J]}$, has been suppressed so far in our presentation.    
  For the scattering solutions, $\Psi_{[J]}$ does indeed depend on the incoming momenta of particles, because the finite volume wave functions, $\Phi_{[J]}$ and $X^{(ij)}_{[J]}$, are constructed from $\Psi_{[J]}$,   momenta dependence remains in finite volume wave functions as well. 
     Therefore, using the infinite volume scattering solutions, $\Psi_{[J]}$, as inputs,    the finite volume quantization conditions, Eq.(\ref{rddet}),  yield discrete energy spectra as the consequence of periodic cubic lattice structure. In other words, the discrete energy spectra in finite volume is the result of long distance correlation effects of particles in a periodic lattice structure. Meanwhile, the specific patterns of discrete spectra rely on the short distance interaction that is described by scattering amplitudes or "amplitudes carrier" wave functions, $\Psi_{[J]}$, in infinite volume. Ultimately, the quantization conditions play the role of imposing constraints on energy spectra due to periodic boundary conditions. The scattering information in infinite volume and periodic boundary conditions are  combined together by  finite volume wave functions,  $\Phi_{[J]}$ and $X^{(ij)}_{[J]}$.    
 
The secular equations, Eq.~(\ref{rdseceq}), and corresponding quantization conditions, Eq.~(\ref{rddet}), can be further reduced by removing the center-of-mass motion. However, the choice of relative coordinates for the $N$-body system normally can be made quite arbitrary. In Section \ref{relmotion},  the specific choice  \cite{Faddeev:1993}  is described by forming the succession of subsystems of $N$ particles in such a way that the subsystems are obtained by successive joining of particle-$2$, particle-$3$, $\cdots$, particle-$N$  to particle-$1$, {\it i.e.} $(12)$, $((12)3)$, $(((12)3)4)$, $\cdots$.

\subsection{Removal of center of mass motion}\label{relmotion}

Since in this work, we are mainly interested in scattering solutions, the set of incoming  particle's momenta is also introduced to label the $N$-body wave functions:  $\{ \mathbf{p}\} =\{ \mathbf{p}_1, \mathbf{p}_2, \cdots, \mathbf{p}_N\}$, where $\mathbf{p}_i$ stands for the  momentum of the $i$-th incoming particle. The total energy of  $N$ particles is given by $E=  \sum_{i=1}^{N} p^{2}_{i}/2 m$.
The center-of-mass motion and internal motion of the  $N$-body system can be separated out by changing variables of coordinates system. One particular choice of  the relative coordinates and momenta \cite{Faddeev:1993} may be given respectively by 
\begin{align}
&\bm{\rho}_{(12), n} = \sqrt{\frac{2n}{n+1}} \left ( \frac{1}{n} \sum_{k=1}^{n}  \mathbf{ x}_{k} - \mathbf{ x}_{n+1} \right ),   \nonumber \\
& \mathbf{ q}_{(12), n} = \sqrt{\frac{n}{2(n+1)}} \left ( \frac{1}{n} \sum_{k=1}^{n}  \mathbf{ p}_{k} - \mathbf{ p}_{n+1} \right ),   \label{relrqcord}
\end{align}
where \mbox{$n<N$}.  The quantities  $ \bm{ \rho}_{(12), n}$ and $ \mathbf{ q}_{(12), n}$ may be interpreted as the relative coordinate and momentum between the \mbox{$(n+1)$}-th particle and the center of  mass coordinate or momentum of the cluster of particles  $(1,2,\cdots, n)$, respectively. 
 The index $(12)$ is used to label the special choice we made on  relative coordinates and momenta, so that  for \mbox{$n=1$}, the relative coordinate and momenta are defined between  particle-1 and particle-2,
\begin{equation}
 \bm{ \rho}_{(12), 1} =  \mathbf{ x}_{1} - \mathbf{ x}_{2}  , \ \ \mathbf{ q}_{(12), 1} =  \frac{1}{2} \left ( \mathbf{ p}_{1} - \mathbf{ p}_{2} \right ).
\end{equation}
 The    complete sets of relative coordinates and momenta  are given by  \mbox{$ \{ \bm{ \rho}_{(12)}  \} = \{\bm{ \rho}_{(12),1}  , \bm{ \rho}_{(12),2}, \cdots , \bm{ \rho}_{(12),N-1}    \}$} and  \mbox{$ \{ \mathbf{ q}_{(12)}  \} = \{\mathbf{ q}_{(12),1}  , \mathbf{ q}_{(12),2}, \cdots , \mathbf{ q}_{(12),N-1}    \}$}   respectively.
The center of mass coordinate and momentum of the $N$-body system are  
\begin{equation}
\mathbf{ R} = \frac{1}{N}  \sum_{i=1}^{N}  \mathbf{ x}_{i}, \ \ \ \  \mathbf{ P} = \sum_{i=1}^{N}  \mathbf{ p}_{i}.
\end{equation}
The total energy of the  $N$-particle  system  is given by \mbox{$E=  \sum_{n=1}^{N-1} \mathbf{ q}^{2}_{(12),n}/m + \mathbf{ P}^{2}/2mN $}, and the kinetic energy operator of the $N$-body system is
\begin{equation}
\hat{T} =-  \sum_{n=1}^{N-1}  \frac{ \nabla^{2}_{\bm{ \rho}_{(12),n}}  }{m} - \frac{1}{N}  \frac{\nabla^{2}_{\mathbf{ R}} }{2m}.
 \end{equation}
 
  In terms of the set $\{ \bm{\rho}_{(12)}\}$, the infinite-volume $N$-body wave function is given by \mbox{$ \Psi( \{ \mathbf{ x}  \} ,  \{ \mathbf{ p}  \}) = e^{i \mathbf{ P} \cdot \mathbf{ R}}  \psi( \{ \bm{ \rho}_{(12)}  \} ,  \{ \mathbf{ q}_{(12)}  \})$}, where the center of mass motion is represented by a plane wave, $ e^{i \mathbf{ P} \cdot \mathbf{ R}}$, and the wave function $\psi$ describes the internal motions of the $N$-body system. The same applies to the finite-volume wave function. 
   The $N$-body wave function representing the relative motion in  finite volume is thus given  by 
  \begin{align}
  &  \phi( \{ \bm{ \rho}_{(12)}  \} ,  \{ \mathbf{ q}_{(12)}  \})    \nonumber \\
  &   = \frac{1}{\mathcal{N}} \sum_{ \{ \mathbf{ n}_{\mathbf{ x}} \} \in \mathbb{Z}^{3}} e^{i \frac{ \mathbf{ P} }{N} \cdot \sum_{i=1}^{N} \mathbf{ n}_{\mathbf{ x}_{i}}   }       \psi  (\{ \ddot{\bm{ \rho}}_{(12)}   \}, \{ \mathbf{ q}_{(12)} \})  ,  \label{phiconstr}
 \end{align} 
 where the $n$-th element of the set $\{ \ddot{\bm{ \rho}}_{(12)}  \}$ is 
 \begin{equation}
\ddot{ \bm{ \rho}}_{(12),n}  = \bm{ \rho}_{(12),n}  + \sqrt{\frac{2n}{n+1}} \left ( \frac{1}{n} \sum_{k=1}^{n}  \mathbf{ n}_{ \mathbf{ x}_{k}  }-  \mathbf{ n}_{\mathbf{ x}_{n+1} } \right ) L. 
 \end{equation}
In order to remove one redundant element of the set $\{ \mathbf{ n}_{\mathbf{ x}} \} $, a  subset $ \{ \mathbf{ n}_{(12)} \} $  is introduced, 
 \begin{equation}
 \{ \mathbf{ n}_{(12)} \} = \{ \mathbf{ n}_{(12),1},  \mathbf{ n}_{(12),2},   \cdots,  \mathbf{ n}_{(12),N-1} \}   \in \mathbb{Z}^{3} ,
\end{equation}
 where    the $k$-th element is given by \mbox{$\mathbf{ n}_{(12),k} = \mathbf{ n}_{\mathbf{ x}_{k }} -\mathbf{ n}_{\mathbf{ x}_{k+1}}$}. With the introduction of the set $ \{ \mathbf{ n}_{(12)} \} $ in this particular way, all the elements of set $\{ \mathbf{ n}_{(12)} \} $  still belong  in  $\mathbb{Z}^{3}$. We also  find
   \begin{equation}
\ddot{ \bm{ \rho}}_{(12),n}  = \bm{ \rho}_{(12),n}  + \sqrt{\frac{2}{n(n+1)}}   \sum_{k=1}^{n} k  \mathbf{ n}_{ (12),k   } L, 
 \end{equation}
 and
 \begin{equation}
 \sum_{i=1}^{N}  \mathbf{ n}_{ \mathbf{ x}_{i}  } = \sum_{k=1}^{N-1} k  \mathbf{ n}_{ (12),k   } L  + N  \mathbf{ n}_{ \mathbf{ x}_{N}  }.
 \end{equation}
 Hence, after removal  of the center-of-mass motion,  the finite-volume relative wave function, $\phi$, is related to $\psi$ by
\begin{align}  
& \phi    ( \{ \bm{ \rho }_{(12) }   \},  \{ \mathbf{ q}_{(12)} \})  \nonumber \\
&  =    \sum_{ \{ \mathbf{ n}_{(12)} \}  \in \mathbb{Z}^{3} }   e^{i \frac{\mathbf{ P}}{N} \cdot    \left ( \sum_{k=1}^{N-1 } k \mathbf{ n}_{(12),k}     \right ) L  }   \psi   ( \{\ddot{ \bm{ \rho}}_{(12)}  \}, \{ \mathbf{ q}_{(12)} \}).
\end{align}
 and $\phi  $ satisfies periodic boundary condition,
\begin{align}
& \phi   (\{ \ddot{\bm{ \rho}}_{(12)}  \}, \{ \mathbf{ q}_{(12)} \})  \nonumber \\
& \quad \quad =      e^{-i \frac{\mathbf{ P}}{N} \cdot    \left ( \sum_{k=1}^{N-1 } k \mathbf{ n}_{(12),k}     \right ) L  }   \phi  (\{ \bm{ \rho}_{(12) }   \}, \{ \mathbf{ q}_{(12)} \}) .
\end{align}
 
 Clearly, the   set   $\{ \bm{ \rho}_{(12)} \}$  cannot be the only choice of independent relative coordinates. By exchanging the labels of particle, such as $1 \leftrightarrow i$ and $2 \leftrightarrow j$  in  set $\{ \bm{ \rho}_{(12)} \}$,     another  independent coordinate set,   $\{ \bm{ \rho}_{(ij)} \}$, may be obtained. For example,  relabeling $2 \leftrightarrow 3$ in $\{ \bm{ \rho}_{(12)} \}$,  the elements of set $\{ \bm{\rho}_{(13)} \}$  are given   by
 \begin{align}
 & \bm{ \rho}_{(13), 1} =  \mathbf{ x}_{1}-  \mathbf{ x}_{3} = \frac{\bm{ \rho}_{(12), 1}   + \sqrt{3}  \bm{ \rho}_{(12), 2}  }{2} , \nonumber \\
 &   \bm{ \rho}_{(13), 2}  =\frac{2}{\sqrt{3}} \left ( \frac{ \mathbf{ x}_{1}+  \mathbf{ x}_{3} }{2} - \mathbf{ x}_{2}   \right ) = \frac{  \sqrt{3}  \bm{ \rho}_{(12), 1}   - \bm{ \rho}_{(12), 2}  }{2} , \nonumber \\
 &\bm{ \rho}_{(13), n} = \bm{ \rho}_{(12), n}   , \ \ n >2.
 \end{align}
 The different sets $\{ \bm{ \rho}_{(ij)} \}$ and $\{ \bm{ \rho}_{(i'j')} \}$  are linked by a linear transformation,
\begin{equation}
\{ \bm{ \rho}_{(ij)} \} = \Gamma^{(ij),(i'j')}    \{ \bm{ \rho}_{(i'j')} \},
\end{equation} 
for example,
\begin{align}
\Gamma^{(13),(12)} = \begin{bmatrix}  \frac{1}{2} & \frac{\sqrt{3}}{2} & 0 \\ \frac{\sqrt{3}}{2} & - \frac{1}{2} &  0 \\  0 & 0 & \mathbb{I} \end{bmatrix}  .
\end{align}
Because of the way of construction, the transfer matrix $\Gamma$ between two different sets is an orthogonal matrix, $  \Gamma^{T}    \Gamma   = \mathbb{I}$. Hence,
both the sum, $\sum_{n=1}^{N-1} \bm{ \rho}_{(ij),n}^{2} $, and the $N$-body volume element, $\int \prod_{n=1}^{N-1} d \bm{ \rho}_{(ij),n}$  are  invariant under 
 this transformation of sets. Similarly, both total energy and kinetic energy operator are also  invariant under transformation between  sets of relative coordinates and momenta.

 With introduction of these sets of relative coordinates,  the finite volume $N$-body wave function, $\phi_{[J]}$,   may be written as, 
\begin{equation}
  \phi_{[J]}( \{ \bm{ \rho}_{(12)}  \} ,  \{ q_{(12)}  \}) =  \sum^{N}_{(i  < j)=1}  \phi^{(ij)}_{[J]}( \{ \bm{ \rho}_{(ij)}  \} ,  \{ q_{(12)}  \}), \label{phiJ}
\end{equation}
where $\phi^{(ij)}_{[J]}$ satisfies
\begin{align}
& \left ( \sigma^{2} - \hat{T}_{\bm{\rho}}       \right )   \phi^{(ij)}_{[J]}( \{ \bm{ \rho}_{(ij)}  \} ,  \{ q_{(12)}  \}) \nonumber \\
& \quad \quad  \quad \quad     =  m V^{(L)}_{(ij)} (\bm{ \rho}_{(ij),1})   \phi_{[J]} ( \{ \bm{ \rho}_{(12)}  \} ,  \{ q_{(12)}  \})  ,\label{fvrelwavschr}
\end{align}
 and we introduced rescaled total energy and kinetic energy operators, $\sigma^{2} =  m E - \mathbf{ P}^{2}/2 N$ and $\hat{T}_{\bm{\rho}} = -\sum_{n=1}^{N-1}  \nabla^{2}_{\bm{ \rho}_{(ij), n}}  $. The construction of  $\phi^{(ij)}_{[J]}$ is given by
 \begin{align}  
\phi^{(ij)}_{[J]}    ( \{ \bm{ \rho}_{(ij) }   \}, & \{ q_{(12)} \})  =    \sum_{ \{ \mathbf{ n}_{(ij)} \}  \in \mathbb{Z}^{3} }   e^{i \frac{\mathbf{ P}}{N} \cdot    \left ( \sum_{n=1}^{N-1 } n \mathbf{ n}_{(ij),n}     \right ) L  }  \nonumber \\
& \quad   \times  \psi^{(ij)}_{[J]}   ( \{ \ddot{\bm{ \rho}}_{(ij)}   \}, \{ q_{(12)} \}), \label{fvrefwavconstr}
\end{align}
where     $ \{ \mathbf{ n}_{(ij)} \} =  \{ \mathbf{ n}_{(ij),1} , \mathbf{ n}_{(ij),2} , \cdots , \mathbf{ n}_{(ij),N-1} \}  $, and  the set  $ \{ \mathbf{ n}_{(ij)} \}  $  may be related to  $ \{ \mathbf{ n}_{(12)} \} $ by relabeling $1 \leftrightarrow i$ and $2 \leftrightarrow j$.  The set $ \{ \ddot{\bm{ \rho}}_{(ij)}   \}$ is defined by 
\begin{equation}
 \ddot{ \bm{ \rho}}_{(ij),n}  = \bm{ \rho}_{(ij),n}  + \sqrt{\frac{2}{n(n+1)}}   \sum_{k=1}^{n} k  \mathbf{ n}_{ (ij),k   } L, \  \ (  \mathbf{ n}_{ (ij),k   } \in \mathbb{Z}^{3}).
\end{equation}
 The wave functions, $ \psi^{(ij)}_{[J]}$, are the solutions of  infinite-volume Schr\"odinger equations,
 \begin{align}
&\left ( \sigma^{2} - \hat{T}_{\bm{\rho}}       \right )    \psi^{(ij)}_{[J]}( \{ \bm{ \rho}_{(ij)}  \} ,  \{ q_{(12)}  \}) \nonumber \\
& \quad \quad  \quad \quad      =  m V_{(ij)} (\bm{ \rho}_{(ij),1})   \psi_{[J]} ( \{ \bm{ \rho }_{(12)}  \} ,  \{ q_{(12)}  \})  , \label{freerelwavschr}
\end{align}
and the total  infinite-volume $N$-body wave function is given by, $\psi_{[J]} = \psi^{(0)}_{[J]} +\sum^{N}_{(i  < j)=1}  \psi^{(ij)}_{[J]}$,  where $ \psi^{(0)}_{[J]}$ refers to the incoming free wave.

  After removal of the center-of-mass motion, the determinant conditions    are now given by
 \begin{equation}
 \det \left [  \langle \phi_{[J']}    |  \hat{V}_{(ij)}    \left (   |   \chi^{(ij)}_{[J]}   \rangle  - | \phi_{[J]} \rangle   \right ) \right ]   =0   , \label{rdreldetcond}
\end{equation} 
where    
   \begin{align}
  & \chi^{(ij)}_{[J]}  (\{ \bm{ \rho}_{(12)} \}, \{ q_{(12)} \})  = \sum_{ \overline{\{ \mathbf{ n}_{ (ij)} \}} \in \mathbb{Z}^{3}}     e^{i \frac{\mathbf{ P}}{N} \cdot    \left ( \sum_{n=2}^{N-1 } n \mathbf{ n}_{(ij),n}     \right ) L  }   \nonumber \\
 & \quad \quad \times  \psi_{[J]}  ( \Gamma^{(12),(ij)}  \{  \overline{ \ddot{\bm{\rho}}_{(ij)} } \} , \{ q_{(12)} \})  .   \label{overlinephiijpwa}
 \end{align} 
The set $\overline{\{ \mathbf{ n}_{(ij)} \}}  $ refers to   $  \{ \mathbf{ n}_{(ij),2} , \mathbf{ n}_{(ij),3} , \cdots , \mathbf{ n}_{(ij),N-1} \}  $ which is a subset of $\{ \mathbf{ n}_{(ij)} \} $ by excluding element $ \mathbf{ n}_{(ij),1}$.    The  set $ \{  \overline{ \ddot{\bm{\rho}}_{(ij)} }\} $ is defined by
  \begin{equation}
\overline{\ddot{ \bm{ \rho}}_{(ij),n}  } = \bm{ \rho}_{(ij),n}  + \sqrt{\frac{2}{n(n+1)}}   \sum_{k=2}^{n} k  \mathbf{ n}_{ (ij),k   } L, \   \  (\mathbf{ n}_{ (ij),k   } \in \mathbb{Z}^{3}).
 \end{equation}

Noting   the  invariance of set $\{ n_{(ij)} \}$ we are able to write $ \phi_{[J]} $ in a similar way  as Eq.~(\ref{overlinephiijpwa}),
    \begin{align}
  & \phi_{[J]}  (\{ \bm{ \rho}_{(12)} \}, \{ q_{(12)} \})  = \sum_{  \{ \mathbf{ n}_{ (ij)} \} \in \mathbb{Z}^{3}}     e^{i \frac{\mathbf{ P}}{N} \cdot    \left ( \sum_{n=1}^{N-1 } n \mathbf{ n}_{(ij),n}     \right ) L  }   \nonumber \\
 & \quad \quad \times  \psi_{[J]}  ( \Gamma^{(12),(ij)}  \{    \ddot{\bm{\rho}}_{(ij)}  \} , \{ q_{(12)} \})  ,
 \end{align} 
 which is equivalent to Eq.~(\ref{phiconstr}). Hence, the relation between $\phi_{[J]}$ and $\chi^{(ij)}_{[J]} $ can be made more clear,
   \begin{align}
  & \phi_{[J]}  (\{ \bm{ \rho}_{(12)} \}, \{ q_{(12)} \})  = \sum_{  \mathbf{ n}_{(ij),1} \in \mathbb{Z}^{3}}     e^{i \frac{\mathbf{ P}}{N} \cdot      \mathbf{ n}_{(ij),1}       L  }   \nonumber \\
 & \quad \quad \times  \chi^{(ij)}_{[J]}  ( \Gamma^{(12),(ij)}  \{  \widetilde{ \bm{ \rho}}_{(ij)}  \} , \{ q_{(12)} \})    , \label{phichiij}
 \end{align} 
where the $n$-th element of  $\{  \widetilde{ \bm{ \rho}}_{(ij)}  \}$  is given by
  \begin{equation}
 \widetilde{ \bm{ \rho}}_{(ij),n}  = \bm{ \rho}_{(ij),n}  +   \mathbf{ n}_{ (ij),1   } L, \  \  \   (  \mathbf{ n}_{ (ij),1   } \in \mathbb{Z}^{3} ),
 \end{equation}
 and $ \{ \ddot{ \bm{ \rho}}_{(ij)} \}  = \{ \overline{\ddot{ \bm{ \rho}}_{(ij)}  } \} + \{  \widetilde{ \bm{ \rho}}_{(ij)}  \}  $.

\subsection{Relation to two-body L\"uscher formula}\label{reltoLuscher}

 In the simplest case, $N=2$, there is only one  relative coordinate and relative momentum so that the particle index can be 
  dropped and in  what follows they are 
   denoted by $\mathbf{ r}$ and $\mathbf{ q}$, respectively. Notice that $\chi^{(ij)}_{[J]}$ is now reduced to $\psi_{[J]}$ for the two-body system. Therefore, the 
determinant condition in Eq.~(\ref{rdreldetcond}) is  given by
   \begin{equation}
 \det \left [  \langle \phi_{[J']}    |  \hat{V}   \left (     | \psi_{[J]}   \rangle  - | \phi_{[J]} \rangle   \right )   \right ]   =0   ,  \label{2bdetcond}
\end{equation} 
where $[J]=JM$ stand for quantum numbers of a specific partial wave, and
 \begin{equation}  
\phi_{[J]}    ( \mathbf{ r}   , q  )  =    \sum_{  \mathbf{ n}  \in \mathbb{Z}^{3} }   e^{i \frac{\mathbf{ P}}{2} \cdot     \mathbf{ n}    L  }    \psi_{[J]}   (  \mathbf{ r}  +   \mathbf{ n}     L ,  q  ).
\end{equation}
Assuming the potential is of finite range, \mbox{$V(r)=0$} for \mbox{$R<r<L$},  in  the region outside of the potential the  infinite-volume wave function has the form,
\begin{align}
 & \psi_{[J]} (\mathbf{ r},q) \stackrel{r>R}{ =}  \psi^{(out)}_{[J]} (\mathbf{ r},q) ,\nonumber \\
 & \psi^{(out)}_{[J]} (\mathbf{ r},q)  = (4\pi) i^{J} Y_{[J]}  (   \mathbf{ r}  ) \left [ j_{J} (q r) + i t_{J} (q) h_{J}^{(+)} (q r) \right],
\end{align}
where   $t_{J}$ is the partial-wave two-body scattering amplitude, and $t_{J}$ is normalized by the unitarity relation, Im~$ \left [ 1/ t_{J} \right ]=-1$.  Inside the potential region, the wave function is given by the solution of the Schr\"odinger equation, and is denoted by $\psi^{(in)}_{[J]} (\mathbf{ r},q) $ from now on. The matrix element of the determinant condition, Eq.~(\ref{2bdetcond}), is given by
\begin{align}
\int_{r<R} d \mathbf{ r} \phi^{*}_{[J']} (\mathbf{ r},q) V(r) \left [ \psi^{(in)}_{[J]} (\mathbf{ r},q) - \phi_{[J]} (\mathbf{ r},q)  \right ],
\end{align}
Using the fact that
 \begin{align}  
& \phi_{[J]}    ( \mathbf{ r}   ,  q  )  \stackrel{r<R}{ =}\psi^{(in)}_{[J]} (\mathbf{ r},q)  - \psi^{(out)}_{[J]} (\mathbf{ r},q)    \nonumber \\
& +  (4\pi) i^{J} i t_{J} (q) \sum_{  \mathbf{ n}  \in \mathbb{Z}^{3} }   e^{i \frac{\mathbf{ P}}{2} \cdot     \mathbf{ n}    L  }      Y_{[J]}  (  \mathbf{ r}  +   \mathbf{ n}     L )   h_{J}^{(+)} (q |\mathbf{ r}  +   \mathbf{ n}     L |) ,
\end{align}
and the relation 
\begin{align}
&  \sum_{  \mathbf{ n}  \in \mathbb{Z}^{3} }   e^{i \frac{\mathbf{ P}}{2} \cdot     \mathbf{ n}    L  }      Y_{[J]}  (  \mathbf{ r}  +   \mathbf{ n}     L )   h_{J}^{(+)} (q |\mathbf{ r}  +   \mathbf{ n}     L |)   \nonumber \\
&= \sum_{[j]} Y_{[j]} (\mathbf{ r}) \left [ \delta_{[J], [j]} i n_{J} (qr) - i \mathcal{M}_{[J] ,[j]}^{(\frac{\mathbf{ P}}{2})}  (q) j_{j} (q r)\right ]  , \label{luscherlatsum}
\end{align}
derived in Eq.~(B1) and Eq.~(B3) in 
 Ref.~\cite{Guo:2012hv}, for expansion coefficient function of the finite volume two-body Green's function,  $\mathcal{M}_{[J] ,[j]}^{(\frac{\mathbf{ P}}{2})}$,     we find
\begin{align}
&   \langle \phi_{[J']}    |  \hat{V}   \left (     | \psi_{[J]}   \rangle  - | \phi_{[J]} \rangle   \right )    \nonumber \\
&=  \sum_{[j]} \left [   \int_{r<R} d \mathbf{ r} \phi^{*}_{[J']} (\mathbf{ r},q) V(r)    j_{j} (q r) Y_{[j]} (\mathbf{ r}) \right ]  \nonumber \\
& \times (4\pi) i^{J}  \left [  \delta_{[J],[j]} \left (1+ i t_{J} (q)  \right)  - t_{J} (q)  \mathcal{M}_{[J], [j]}^{(\frac{\mathbf{ P}}{2})}  (q)\right ] .
\end{align}
Therefore, the determinant condition for the two-body problem, Eq.~(\ref{2bdetcond}),  yields  L\"uscher's  formula,
\begin{align}
\det \left [  \delta_{[J],[j]} \left (\frac{1}{  t_{J} (q) } + i \right)  - \mathcal{M}_{[J], [j]}^{(\frac{\mathbf{ P}}{2})}  (q)\right ] =0.
\end{align}
Hence, the variational approach to the finite -volume few-body system  is  consistent with the L\"uscher approach.

\subsection{Effective two-body formalism}\label{Nbeff}
In this section, we would like to show that the $N$-body    quantization conditions may be recast in a similar form as  the two-body quantization condition  in Eq.~(\ref{2bdetcond}). 

Before proceeding to $N$-body interaction, first of all, let us rewrite the two-body quantization condition of Eq.~(\ref{2bdetcond}) by using the periodicity of the finite-volume wave function. We find
\begin{equation}
\phi^{*}_{[J']}(\mathbf{ r}, q)\phi_{[J]}(\mathbf{ r},q ) =  \sum_{\mathbf{ n} \in \mathbb{Z}^{3}}  \phi^{*}_{[J']}(\mathbf{ r} + \mathbf{ n} L, q)    \psi_{[J]}(\mathbf{ r} + \mathbf{ n} L, q).
\end{equation}
Therefore, the two-body quantization condition, Eq.~(\ref{2bdetcond}), may be expressed as
\begin{align}
& \det \Bigg [ \int d \mathbf{ r} V(r) \bigg [\Omega_{[J'], [J]}  (\mathbf{ r},q)  \nonumber \\
&\quad \quad \quad \quad\quad \quad \quad -  \sum_{\mathbf{ n} \in \mathbb{Z}^{3}} \Omega_{[J'], [J]}  (\mathbf{ r} + \mathbf{ n} L,q) \bigg ]  \Bigg ]=0, \label{2beff}
\end{align}
where $\Omega_{[J'], [J]}  (\mathbf{ r},q)=\phi^{*}_{[J']}(\mathbf{ r}  , q)    \psi_{[J]}(\mathbf{ r}  , q) $.

Next, for the $N$-body problem,  we revisit the determinant condition in channel $(ij)$ in a   explicit format,
\begin{align}
&  \det \Bigg [ \int \prod_{n=1}^{N-1} d \bm{\rho}_{(ij),n} \phi^{*}_{[J']} ( \{ \bm{\rho}_{(12)} \}, \{q_{(12)} \}) m V_{(ij)}( \bm{\rho}_{(ij),1}) \nonumber \\
& \times \left [\chi^{(ij)}_{[J]} ( \{  \bm{\rho}_{(12)} \}, \{q_{(12)} \}) -\phi_{[J]} ( \{ \bm{\rho}_{(12)} \}, \{q_{(12)} \})  \right ]  \Bigg ] =0, \label{seculareqij2b}
\end{align}
where $ \chi^{(ij)}_{[J]}$ and $\phi_{[J]} $  are given by Eq.~(\ref{overlinephiijpwa}) and Eq.~(\ref{phichiij}), respectively.
 Because of the relation between $ \chi^{(ij)}_{[J]}$ and $\phi_{[J]} $ given in Eq.~(\ref{phichiij}), similarly to Eq.~(\ref{2beff}), it is  advantageous  to introduce the  quantity  $\Omega^{(ij)} (\bm{\rho}_{(ij),1})$ 
again by integrating out all the relative coordinates  except for $\bm{\rho}_{(ij),1}$,
 \begin{align}
& \Omega^{(ij)}_{[J'],[J]} (\bm{\rho}_{(ij),1} , \{q_{(12)} \} )=\int \prod_{n=2}^{N-1} d \bm{\rho}_{(ij),n} \nonumber \\
 & \quad \times \phi^{*}_{[J']} ( \{ \bm{\rho}_{(12)} \}, \{q_{(12)} \}) \chi^{(ij)}_{[J]} ( \{  \bm{\rho}_{(12)} \}, \{q_{(12)} \}).
 \end{align}
Using the periodicity of the wave function $\phi$,    the quantization conditions, Eq.~(\ref{seculareqij2b}), now have a form  that resembles the two-body quantization condition given in Eq.~(\ref{2beff}),
 \begin{align}
 & \det  \Bigg\{ \int d \bm{\rho}_{(ij),1}  V_{(ij)}( \bm{\rho}_{(ij),1})    \bigg [ \Omega^{(ij)}_{[J'],[J]} (\bm{\rho}_{(ij),1} , \{q_{(12)} \} ) \nonumber \\
 & \quad \quad  - \sum_{  \mathbf{ n}  \in \mathbb{Z}^{3}}    \Omega^{(ij)}_{[J'],[J]} (\bm{\rho}_{(ij),1}+\mathbf{ n}   L  , \{q_{(12)} \} )\bigg ] \Bigg \}=0. \label{Nbeffsecular}
 \end{align}
 When all the interactions $V_{(i'j')}$ except the interaction between the $i$-th and $j$-th particle are turned off, clearly,  the quantization condition in the $(ij)$ channel in Eq.~(\ref{Nbeffsecular}) is thus reduced to the two-body quantization condition given in Eq.~(\ref{2beff}).

\section{Discussion and conclusion}\label{discussion}

\subsection{Resemblance to isobar model}\label{isobar}
In the past, isobar models \cite{Khuri:1960zz,Bronzan:1963xn,Aitchison:1965kt,Aitchison:1965zz,Aitchison:1966kt,Pasquier:1968zz,Pasquier:1969dt,Guo:2014vya,Guo:2014mpp,Danilkin:2014cra,Guo:2015kla,Guo:2015zqa,Guo:2016wsi, Mai:2017vot}  have been a useful tool to describe few-body interactions, in which the few-body interaction is treated  by taking into account all possible recombinations of two-body subsystems. The two-body subsystems are considered as the dominant contribution compared to three-body force,  
and the few-body interaction correction to the two-body subsystem  is generated by rescattering  between all possible pairs. 
In order to show the similarity of this 
 approach to the isobar  formulation \cite{Khuri:1960zz,Bronzan:1963xn,Aitchison:1965kt,Aitchison:1965zz,Aitchison:1966kt,Pasquier:1968zz,Pasquier:1969dt,Guo:2014vya,Guo:2014mpp,Danilkin:2014cra,Guo:2015kla,Guo:2015zqa,Guo:2016wsi, Mai:2017vot},  we consider a special case, {\it i.e.}, a three-body system with two light spinless particles and one infinitely heavy spinless particle stationed at the origin. The heavy particle is labeled as third particle. The system may be described by  
\begin{align}
& \left [ E+ \frac{1}{2m} \sum_{i=1}^{2} \nabla^{2}_{\mathbf{ r}_{i}}  - \sum_{i=1}^{2} V^{(L)} (r_{i}) + U^{(L)}(\mathbf{ r}_{1} - \mathbf{ r}_{2})  \right ]  \nonumber \\
& \quad \quad \quad \quad \times \phi(\mathbf{ r}_{1}, \mathbf{ r}_{2}; \mathbf{ q}_{1}, \mathbf{ q}_{2}) =0,
\end{align}
where $V^{(L)}$ represents the interactions between  the heavy particle and one of the light particles, and $U^{(L)}$ stands for the interaction between the two light particles. Again, we use the superscript $(L)$ to identify the periodic potential in finite volume. 
We also assume that $U^{(L)}$ is a weak interaction, so that the interaction between the two light particles  is treated as a perturbation, which serves the purpose of discussion in this work. Therefore, the corresponding infinite-volume wave function must have the form of
\begin{equation}
 \psi(\mathbf{ r}_{1}, \mathbf{ r}_{2}; \mathbf{ q}_{1}, \mathbf{ q}_{2})  = \psi(\mathbf{ r}_{1} ; \mathbf{ q}_{1}) \psi(\mathbf{ r}_{2} ; \mathbf{ q}_{2})   + \delta \psi(\mathbf{ r}_{1}, \mathbf{ r}_{2}; \mathbf{ q}_{1}, \mathbf{ q}_{2}) ,
\end{equation}
where the first term is the solution of the system with zero interaction between two lights particles. 
The second term, $\delta \psi$, can be considered as perturbative contribution when the weak $U$-potential is turned on. The two-body  infinite-volume wave function, $\psi(\mathbf{ r}_{i}; \mathbf{ q}_{i})$, is given by the solution of
\begin{equation}
\left [ \frac{q_{i}^{2}+\nabla^{2}_{\mathbf{ r}_{i}} }{2m}    -  V (r_{i})    \right ]  \psi(\mathbf{ r}_{i}; \mathbf{ q}_{i})=0,
\end{equation}
with  $E = q_{1}^{2}/2m + q_{2}^{2}/2m$. Following the argument provided in previous sections, the finite-volume wave function is constructed from the infinite-volume wave function. The two ingredients $\phi$ and $\chi^{(1/2)}$ of the secular equations thus also have the forms,
\begin{equation}
\phi(\mathbf{ r}_{1}, \mathbf{ r}_{2}; \mathbf{ q}_{1}, \mathbf{ q}_{2})  = \phi(\mathbf{ r}_{1} ; \mathbf{ q}_{1}) \phi(\mathbf{ r}_{2} ; \mathbf{ q}_{2})   + \delta \phi(\mathbf{ r}_{1}, \mathbf{ r}_{2}; \mathbf{ q}_{1}, \mathbf{ q}_{2}) , 
\end{equation}
and
\begin{align}
& \chi^{(1)}(\mathbf{ r}_{1}, \mathbf{ r}_{2}; \mathbf{ q}_{1}, \mathbf{ q}_{2}) \nonumber \\
 & \quad \quad  = \psi(\mathbf{ r}_{1} ; \mathbf{ q}_{1}) \phi(\mathbf{ r}_{2} ; \mathbf{ q}_{2})   + \delta \chi^{(1)}(\mathbf{ r}_{1}, \mathbf{ r}_{2}; \mathbf{ q}_{1}, \mathbf{ q}_{2}) ,  \nonumber \\
 & \chi^{(2)}(\mathbf{ r}_{1}, \mathbf{ r}_{2}; \mathbf{ q}_{1}, \mathbf{ q}_{2}) \nonumber \\
 & \quad \quad = \phi(\mathbf{ r}_{1} ; \mathbf{ q}_{1}) \psi(\mathbf{ r}_{2} ; \mathbf{ q}_{2})   + \delta \chi^{(2)}(\mathbf{ r}_{1}, \mathbf{ r}_{2}; \mathbf{ q}_{1}, \mathbf{ q}_{2}) , 
\end{align}
where $\phi(\mathbf{ r}_{i}; \mathbf{ q}_{i}) = \sum_{\mathbf{ n}_{i} \in \mathbb{Z}^{3}} \psi(\mathbf{ r}_{i} + \mathbf{ n}_{i} L; \mathbf{ q}_{i})$.  The construction of $\delta \phi$ and $\delta \chi^{(i)}$ can also be obtained based on $\delta \psi$, accordingly, but the specific expressions are not crucial for our brief   discussion. The  two secular equations may also be treated as a perturbation; for example, for channel $(13)$, we obtain
\begin{align}
& \sum_{[J]} \bigg[ \int d \mathbf{ r}_{1} \left ( \int  d \mathbf{ r}_{2}  \phi_{[J']}^{*}(\mathbf{ r}_{1}, \mathbf{ r}_{2}; q_{1}, q_{2}) \phi_{[L_{2}]} (\mathbf{ r}_{2} ; q_{2})  \right )  V(r_{1})   \nonumber \\
& \quad \quad        \times \left [ \psi_{[L_{1}]} (\mathbf{ r}_{1} ; q_{1})  -  \phi_{[L_{1}]} (\mathbf{ r}_{1} ; q_{1})  \right ]  + \delta U_{[J'],[J]} \bigg ] c_{[J]} =0, 
\end{align}
where  $\delta U$ stands for the perturbative contribution to the secular equation from the weak $U$-potential. The three-body quantum numbers set, $[J]$, is constructed from two-body quantum numbers, $[L_{i}]$, by $[J]=[L_{1}] \bigotimes [L_{2}]$. In the limit of $\delta U \rightarrow 0$,   two secular equations yield two independent two-body quantization conditions,
\begin{align}
\det \left [  \delta_{[L_{i}],[l_{i}]} \left (\frac{1}{  t_{L_{i}} (q_{i}) } + i \right)  - \mathcal{M}_{[L_{i}], [l_{i}]}^{(\mathbf{ 0})}  (q_{i})\right ] =0, \ i=1,2.
\end{align}
and both $q_{i}$'s are quantized  independently according to the corresponding L\"uscher  formula. The three-body correction  with weak $U$-potential may be obtained by  perturbation. In the current approach, the physical picture is presented in a way that is similar to the three-body rescattering effect corrected isobar model, \cite{Khuri:1960zz,Bronzan:1963xn,Aitchison:1965kt,Aitchison:1965zz,Aitchison:1966kt,Pasquier:1968zz,Pasquier:1969dt,Guo:2014vya,Guo:2014mpp,Danilkin:2014cra,Guo:2015kla,Guo:2015zqa,Guo:2016wsi}. That is to say that the three-body system considered in this subsection can be treated as two two-body isobar subsystems: $(13)$ and $(23)$, which yield two independent two-body quantization conditions to $q_{1,2}$.   The three-body rescattering correction to isobar subsystems produces an energy shift on quantized two-body energies: $q_1^2/2m +q_2^2/2m + \delta E$.

\subsection{Summary}

In this work, we propose a variational approach to the finite-volume $N$-body problem. In order to fulfill the periodic boundary conditions, the trial wave functions are constructed by linear superposition of all the solutions of the infinite-volume wave functions centered at each image of the periodic cubic boxes, given that the infinite-volume wave functions may be obtained by standard methods. In this 
 approach short-range $N$-body dynamics is    local to each box and long-range 
  correlations correspond to particles 
   travel through the entire periodic structure of the lattice. In  other words, the short-range dynamics are determined by  infinite-volume wave functions, and finite-volume wave functions control the long-range correlations that eventually yield the discrete energy spectra because of the periodic structure of lattice.  
No explicit analytic expressions between discrete lattice eigenvalues and the scattering amplitude, such as present in the two-body L\"uscher formula,  can be given by the variational approach for $N>2$. Instead, the discrete energy spectra and $N$-body scattering amplitudes are linked in a  rather complicated  way. Nevertheless 
 the method is has potential advantages 
  for systems with $N >>2$.   
By combining the variational approach with the Faddeev approach, the $N(N-1)/2$  quantization conditions are obtained ultimately. In the end, these quantization conditions can be expressed in a way that   resemble   rescattering in the isobar approach, The overall rescattering corrections can also be written as a collective effect by integrating out the  of degrees of freedom other then as selected  two-body subsystem.

\section{ACKNOWLEDGMENTS}
  We   acknowledge support from the Department of Physics and Engineering, California State University, Bakersfield, CA.  P.G.   would like to thank Vladimir~Gasparian and David~Gross for numerous fruitful discussions.   This research was supported in part by the National Science Foundation under Grant No. NSF PHY-1748958,  NSF PHY-1415459, and by the U.S. Department of Energy, Office of Science, Office of Nuclear
Physics under contracts 
 no. ~DE-AC05-06OR23177 and DE-FG02-87ER40365. M.D. acknowledges support by the NSF Career grant No. PHY-1452055.

\appendix

\section{Variational principle}\label{variational}
For a complex system, most calculations are based on approximate methods, the variational principle is one of most commonly used approaches. In this section, we briefly outline the main idea of the variational principle as the approximate solution to a general quantum system \cite{Kohn:1948col,Kohn:1954KKR}, which satisfies the Schr\"odinger equation,
\begin{equation}
   \hat{H}     \Psi  \rangle  =E | \Psi  \rangle .
\end{equation}
The  trial wave function $\Psi$ may be expanded in terms of a set of basis functions which satisfy certain boundary conditions or symmetries of the system,
\begin{equation}
| \Psi  \rangle  = \sum_n c_n | n \rangle.
\end{equation}
The   solution of the Schr\"odinger equation is thus given by the variational principle:
\begin{equation}
\frac{\partial \Lambda}{\partial c^*_n} =0,
\end{equation}
where
\begin{equation}
\Lambda = \sum_{n, n'} c^*_n  \langle n | E-\hat{H} | n'\rangle c_{n'}.
\end{equation}
Thus, the variational principle yields secular equations
\begin{equation}
 \sum_{ n'}  \left [ E   \langle n |   n'\rangle - \langle n | \hat{H} | n'\rangle \right  ]  c_{n'} =0,
\label{eq:seculareasy}
\end{equation}
and the non-trivial solutions exist only if
\begin{equation}
  \det \left [ E   \langle n |   n'\rangle - \langle n | \hat{H} | n'\rangle \right  ] =0.
\end{equation}

\section{$N$-body Faddeev equations in infinite volume}\label{faddeeveq}
In this section, for completeness, we give a brief summary of the Faddeev equations for the interaction of $N$ particles in general. The non-relativistic  $N$-body dynamics   is   described by the Schr\"odinger equation,
\begin{equation}
 ( E -\hat{T}      - \hat{V} ) |   \Psi  \rangle  =0.
\end{equation}
Assuming only pair-wise interactions among particles, $\hat{V}=\sum_{(i<j)=1}^N\hat{V}_{(ij)}$, the scattering solution of $N$-body wave function is normally written as the sum of multiple terms \cite{Faddeev:1960su,Faddeev:1965,Gloeckle:1983,Faddeev:1993},
\begin{equation}
| \Psi \rangle  = | \Psi^{(0)}  \rangle  +\sum_{(i<j)=1}^{N} |\Psi^{(ij)}   \rangle ,
\end{equation}
where $\Psi^{(0)} $ stands for the free initial incoming wave of the $N$-particle  state: $( E -\hat{T}     ) |   \Psi^{(0)}  \rangle =0$, and  $\Psi^{(ij)} $ satisfies the equation
  \begin{equation}
   ( E -\hat{T}        ) |   \Psi^{(ij)}  \rangle =  \hat{V}_{(ij)}  |   \Psi   \rangle .  
\end{equation}
In this way, the $N$-body Schr\"odinger equation is turned into $N(N-1)/2$ coupled equations, 
  \begin{equation}
    |   \Psi^{(ij)}  \rangle =  \hat{G}_{(ij)}\hat{V}_{(ij)} \left [  | \Psi^{(0)}  \rangle  +\sum_{(i'<j')=1}^{N; (i'j' \neq ij)} |\Psi^{(i'j')}   \rangle \right ], \label{wavijFaddeev} 
\end{equation}
where the Green's function operator $\hat{G}_{(ij)}$ is given by 
\begin{equation}
\hat{G}_{(ij)} = \left  (E -\hat{T} -\hat{V}_{(ij)} + i \epsilon \right )^{-1}.
\end{equation}
The $\hat{G}_{(ij)}$ is related to the two-body scattering amplitude, $\hat{t}_{ij}$, by
\begin{equation}
\hat{G}_{(ij)} =\hat{G}_{(0)}  \left  ( 1- \hat{t}_{(ij)} \hat{G}_{(0)}  \right ) ,
\end{equation}
where $\hat{G}_{(0)} =   (E -\hat{T}  + i \epsilon  )^{-1}$ stands for free two-body Green's function.

The total $N$-body scattering amplitude  $\hat{T} $ is introduced by
\begin{equation}
\hat{T} |   \Psi^{(0)}  \rangle   = \sum_{(i<j)=1}^N \hat{T}_{(ij)} |   \Psi^{(0)}  \rangle= - \hat{V} |   \Psi  \rangle  ,
\end{equation}
where $\hat{T}_{(ij)} |   \Psi^{(0)}  \rangle  = -\hat{V}_{(ij)} |   \Psi  \rangle  $. Using the relation $\hat{V}_{(ij)}  \hat{G}_{(ij)} = - \hat{t}_{(ij)} \hat{G}_{(0)}  $ and Eq.~(\ref{wavijFaddeev}), we find that the $\hat{T}_{(ij)}$  satisfy the coupled equations
\begin{equation}
\hat{T}_{(ij)} = \hat{t}_{(ij)} - \hat{t}_{(ij)} \hat{G}_{(0)} \sum_{(i'<j')=1}^{N; (i'j' \neq ij)} \hat{T}_{(i'j')}.\label{FaddevTeq}
\end{equation}
The $\hat{T}_{(ij)} $ and $|\Psi^{(ij)}\rangle $ are related in a simple form by
\begin{equation}
|\Psi^{(ij)} \rangle  = - \hat{G}_{(0)} \hat{T}_{(ij)} |   \Psi^{(0)}  \rangle .
\end{equation}
 One of the  advantages of the Faddeev approach is that it demonstrates the   general relations between the subsystem amplitude  and the $N$-body amplitude in a natural way, see Eq.~(\ref{FaddevTeq}), and the two-body amplitude is used as input for $N$-body dynamics. In addition, the unitarity relation of the $N$-body scattering amplitude is also automatically guaranteed by the Faddeev equations.

\section{Solutions of $N$-body problem with $\delta$-function   potentials }
 
To give readers a concrete example of our proposed approach to $N$-body finite volume problem, in this section, we also include a specific example of  short-range interactions between two particles with a $\delta$-function potential.

\subsection{Two-body interaction with $\delta$-function potential }
We consider two-particle scattering  with a pair-wise $\delta$-function potential,   assuming that all  particles are spinless and have equal mass. The bare strength of the $\delta$-function potential between the two  particles is described by  \mbox{$ V_{0}$}.  With the same convention as in Section  \ref{reltoLuscher}, the  infinite-volume wave function is given by the Lippmann-Schwinger equation,
\begin{align}
& \psi_{JM}( \mathbf{ r}  ,  q ) =  ( 4\pi ) i^{J} j_{J} (q r) Y_{JM} (\mathbf{ r})  \nonumber \\
&+ \int d \mathbf{ r'} G_{(0)} ( \mathbf{ r}, \mathbf{ r}'; q )  m V_{0} \delta(\mathbf{ r'}) \psi_{JM}( \mathbf{ r'}  ,  q ) , \label{2blipp}
\end{align}
where $ G_{(0)} $ stands for the free two-body Green's function, 
\begin{align}
& G_{(0)} ( \mathbf{ r}, \mathbf{ r}'; q )  \nonumber \\
&= \int \frac{d \mathbf{ q'}}{(2\pi)^{3}} \frac{ e^{i \mathbf{ q'}  \cdot (\mathbf{ r}-\mathbf{ r'}) } }{q^{2} - \mathbf{ q'}^{2} + i \epsilon} = - \frac{q}{4\pi} i h_{0}^{(+)} ( q |  \mathbf{ r}-\mathbf{ r'}|).
\end{align}
The two-body Lippmann-Schwinger equation, Eq.~(\ref{2blipp}), exhibits an solution,
 \begin{equation}
  \psi_{JM}( \mathbf{ r}  , q )   = ( 4\pi )   i^{J} Y_{JM} (\mathbf{ r})  \left [  j_{J} (q r)  +  \delta_{J,0}   i t_{0} (q) h_{0}^{(+)} (q r) \right] , \label{2bwavfree}
 \end{equation}
 where only $S$-wave contributes to the two-body scattering amplitude, 
 \begin{align}
 t_{0}(q) = - \frac{q}{ \frac{4\pi}{ m V_{0}} + i q h_{0}^{(+)} (q r)|_{r \rightarrow 0}  }.
 \end{align}
 It has been  known   that  singular potentials, such as the $\delta$-function potential,  require in two or higher dimensions   regularization and renormalization \cite{Cavalcanti:1998jx}. Adopting the renormalization scheme proposed in Ref.~\cite{Cavalcanti:1998jx}, a renormalized strength of the $\delta$-function potential, $V_{R}$, is introduced to absorb the divergent part of $h_{0}^{(+)} (q r)|_{r \rightarrow 0} = 1 - \frac{i}{q r}$, and renormalized and bare strengths are related by
 \begin{align}
 \frac{1}{ m V_{0}} =  \frac{1}{ m V_{R}} -    \left.  \frac{1}{4\pi r} \right |_{r\rightarrow 0} . \label{renorm}
 \end{align}
In terms of the renormalized quantity, the $S$-wave two-body scattering amplitude now reads
  \begin{align}
 t_{0}(q) = - \frac{q}{ \frac{4\pi}{ m V_{R}} + i q   },
 \end{align}
and the unitarity relation of the two-body scattering amplitude is guaranteed by ${\rm Im} [t_{0}^{-1}] = -1$.\\
 
 In the finite volume, based on the discussion given in Section  \ref{reltoLuscher}, we find that
 \begin{align}
&   \psi_{JM}   ( \mathbf{ r}   ,  q) - \phi_{JM}   ( \mathbf{ r}   , q) =  (4\pi) i^{J}    \sum_{j m_{j}}  j_{j} (q r) Y_{j m_{j}}(\mathbf{ r}) \nonumber \\
& \times \left [ \delta_{[J], [j]} \left ( 1+ \delta_{J,0} i t_{0} (q) \right) - \delta_{J,0}  t_{0} (q)  \mathcal{M}_{[0] ,[j]}^{(\frac{\mathbf{ P}}{2})} (q) \right ]  .
\end{align} 
The non-trivial solutions of the $\delta$-function potential are given only by $S$-wave, which are determined by
\begin{align}
  \frac{1}{  t_{0} (q) } + i   = \mathcal{M}_{[0], [0]}^{(\frac{\mathbf{ P}}{2})}  (q).
\end{align}

\subsection{ $N$-body interaction with $\delta$-function potential  }\label{Nbwav}

\subsubsection{The solution of $N$-body interaction  in  infinite-volume }\label{Nbwavfree}

Considering again a simple model with short-range $\delta$-function interaction, $V_{(ij)} (\mathbf{ x}_{i}-\mathbf{ x}_{j}) = V_{0} \delta (\bm{ \rho}_{(ij),1})$\cor{,  t}{. T}he solution of the $N$-body interaction may be obtained by a standard Faddeev approach \cite{Faddeev:1960su,Faddeev:1965,Gloeckle:1983,Faddeev:1993}.  For the  scattering with a free $N$-particle incoming wave, the wave function may be expressed as the sum of $1+ N (N-1)/2 $ terms,
\begin{align}
  \psi_{[J]}( \{ \bm{ \rho}_{(12)}  \} ,  & \{  q_{(12)}  \})=  \psi_{[J]}^{(0)}( \{ \bm{ \rho}_{(12)}  \} ,  \{ q_{(12)}  \})    \nonumber \\
 &\ + \sum^{N}_{(i  < j)=1}  \psi_{[J]}^{(ij)}( \{ \bm{ \rho}_{(ij)}  \} ,  \{ q_{(12)}  \}),
\end{align}
where $ \psi^{(0)}_{[J]}$ refers to the incoming free wave, and $ \psi^{(ij)}_{[J]}$ is given by  Lippmann-Schwinger equation,
\begin{align}
 &  \psi_{[J]}^{(ij)} ( \{ \bm{ \rho}_{( ij)}  \} ,  \{ q_{(12)}  \}) \nonumber \\
 & = \int \prod_{n=1}^{N-1} d \bm{ \rho}'_{(ij), n}  G_{(ij)} ( \{ \bm{ \rho}_{(ij)} \}, \{ \bm{ \rho}'_{(ij)} \} ; \sigma)m V_{0}\delta(\bm{ \rho}'_{(ij),1})   \nonumber \\
 & \times    \bigg [ \psi^{(0)}_{[J]}( \{ \bm{ \rho}'_{(12)}  \} ,  \{ q_{(12)}  \})       \nonumber \\
 &   \quad \quad \quad \quad    + \sum^{N; (i'j'\neq ij)}_{(i'   < j' ) =1 }  \psi^{(i' j')}_{[J]} ( \{ \bm{ \rho}'_{( i' j' )}  \} ,  \{ q_{( 12)}  \})   \bigg ]. \label{freewavlipsch}
\end{align}
The Green's function, $G_{(ij)} $, satisfies the equation
\begin{align}
& \left [ \sigma^{2}  - \hat{T}_{\bm{\rho}}   -   m V_{0}\delta(\bm{ \rho}_{(ij),1})    \right ]G_{(ij)} ( \{ \bm{ \rho}_{(ij)} \}, \{ \bm{ \rho}'_{(ij)} \} ; \sigma)    \nonumber \\
&   \quad \quad \quad \quad    \quad \quad   \quad \quad   = \prod_{n=1}^{N-1}  \delta(  \bm{ \rho}_{( ij), n} - \bm{ \rho}'_{( ij), n}  ) , \label{freegreenfunc}
\end{align}
and the solution of Eq.~(\ref{freegreenfunc}) has the form  
\begin{align}
& G_{(ij)} ( \{ \bm{ \rho}_{(ij)} \}, \{ \bm{ \rho}'_{(ij)} \} ; \sigma)   \nonumber \\
& =  \int \prod_{n=1}^{N-1} \frac{ d \mathbf{ q}'_{(ij), n} }{(2\pi)^{3}}   e^{ i \sum_{k=2}^{N-1}   \mathbf{ q}'_{(ij), k} \cdot \left (  \bm{ \rho}_{(ij), k}  -  \bm{ \rho}'_{(ij), k}  \right ) }  \nonumber \\
& \times  \frac{ \psi ( \bm{ \rho}_{(ij),1} , \mathbf{ q}'_{(ij),1} ) \psi^{*}  ( \bm{ \rho }'_{(ij),1} , \mathbf{ q}'_{(ij),1} ) }{\sigma^{2} - \sum_{n=1}^{N-1} \mathbf{ q}^{'2}_{(ij), n}  + i \epsilon },
\end{align}
where  the two-body wave function, $\psi ( \bm{ \rho}_{(ij),1} , \mathbf{ q'}_{(ij),1} )$, is given by 
\begin{align}
& \psi    ( \bm{ \rho}_{(ij),1} , \mathbf{ q}_{(ij),1} )  \nonumber \\
&= e^{ i  \mathbf{ q}_{(ij),1}  \cdot \bm{ \rho}_{(ij),1} } + i t_{0}  ( q_{(ij),1}   )  h_{0}^{(+)} (q_{(ij),1} \rho_{(ij),1}  ). \label{free2bwav}
\end{align}
  Using the expression of the two-body wave function in Eq.~(\ref{free2bwav}),   we find
\begin{align}
& G_{(ij)} ( \{ \bm{\rho}_{(ij)} \}, \{ \bm{ \rho}'_{(ij)} \} ; \sigma)   \nonumber \\
& =  \int \prod_{n=2}^{N-1} \frac{ d \mathbf{ q}'_{(ij), n} }{(2\pi)^{3}}   e^{ i \sum_{k=2}^{N-1}  \mathbf{ q}'_{(ij), k} \cdot \left (  \bm{ \rho}_{(ij), k}  -  \bm{ \rho}'_{(ij), k}  \right ) }  \nonumber \\
& \times  \left [ - \frac{ e^{ i \sqrt{  \sigma^{2} - \sum_{n=2}^{N-1} \mathbf{ q}^{'2}_{(ij), n} }  \left | \bm{ \rho}_{(ij),1} - \bm{ \rho}'_{(ij),1}  \right |}}{4\pi \left | \bm{ \rho}_{(ij),1} - \bm{ \rho}'_{(ij),1}  \right |}  \right. \nonumber \\
& \quad\quad - \frac{1}{4\pi} \frac{  t_{0} (  \sqrt{  \sigma^{2} - \sum_{n=2}^{N-1} \mathbf{ q}^{'2}_{(ij), n} } )  }{\sqrt{  \sigma^{2} - \sum_{n=2}^{N-1} \mathbf{ q}^{'2}_{(ij), n} }} \nonumber \\
& \quad \quad \quad \times  \left. \frac{e^{i \sqrt{  \sigma^{2} - \sum_{n=2}^{N-1} \mathbf{ q}^{'2}_{(ij), n} } \left (\rho_{(ij),1}  + \rho'_{(ij),1}  \right ) }}{  \rho_{(ij),1}  \rho'_{(ij),1}  } \right ] .
\end{align}
With the renormalization relation given in Eq.(\ref{renorm}), we  also find
\begin{align}
& G_{(ij)} ( \{ \bm{\rho}_{(ij)} \}, \{ \bm{ \rho}'_{(ij)} \} ; \sigma)   m V_{0}\delta(\bm{ \rho}'_{(ij),1})     \nonumber \\
& =  \int \prod_{n=2}^{N-1} \frac{ d \mathbf{ q}'_{(ij), n} }{(2\pi)^{3}}   e^{ i \sum_{k=2}^{N-1}  \mathbf{ q}'_{(ij), k} \cdot \left (  \bm{ \rho}_{(ij), k}  -  \bm{ \rho}'_{(ij), k}  \right ) }  \nonumber \\
& \times  i  t_{0}  \left ( \sqrt{  \sigma^{2} - \sum_{n=2}^{N-1} \mathbf{ q}^{'2}_{(ij), n} }  \right  )  \nonumber \\
& \times h_{0}^{(+)} \left ( \sqrt{  \sigma^{2} - \sum_{n=2}^{N-1} \mathbf{ q}^{'2}_{(ij), n} }  \rho_{(ij),1}  \right )   \delta(\bm{ \rho}'_{(ij),1})   . \label{Gij}
\end{align}
 Therefore, we can rewrite $\psi^{(ij)}_{[J]}$ as
\begin{align}
 & \psi^{(ij)}_{[J]}( \{ \bm{ \rho}_{(ij)}  \} ,  \{ q_{(12)}  \})  \nonumber \\
&=    \int \prod_{n=2}^{N-1} \frac{ d \mathbf{ q}'_{(ij), n} }{(2\pi)^{3}}    h_{0}^{(+)} \left ( \sqrt{  \sigma^{2} - \sum_{n=2}^{N-1} \mathbf{ q}^{'2}_{(ij), n} }  \rho_{(ij),1}  \right )    \nonumber \\
& \times    i t_{0}  \left ( \sqrt{  \sigma^{2} - \sum_{n=2}^{N-1} \mathbf{ q}^{'2}_{(ij), n} }  \right )   e^{ i \sum_{k=2}^{N-1}  \mathbf{ q}'_{(ij), k} \cdot    \bm{ \rho}_{(ij), k}  }    \nonumber \\
 & \times \int \prod_{n=1}^{N-1} d \bm{ \rho}'_{(ij), n}   e^{- i \sum_{k=2}^{N-1}  \mathbf{ q}'_{(ij), k} \cdot     \bm{ \rho}'_{(ij), k}  } \delta(\bm{\rho }'_{(ij),1})   \nonumber \\
 & \times   \bigg [ \psi^{(0)}_{[J]}( \{ \bm{ \rho}'_{(12)}  \} ,  \{ q_{(12)}  \})    \nonumber \\
 &   \quad \quad \quad \quad    + \sum^{N; (i'j'\neq ij)}_{(i'   < j' ) =0 }  \psi^{(i' j')}_{[J]} ( \{ \bm{ \rho}'_{( i' j' )}  \} ,  \{q_{( 12)}  \})   \bigg ]. \label{freewav2b}
\end{align}

Next,  we introduce the scattering amplitudes, $T^{(ij)}_{[J]} $, by
\begin{align}
& T^{(ij)}_{[J]}   \left ( \overline{\{  \mathbf{ q}'_{(ij)}  \}} ,  \{  q_{(12)} \} \right  )   \nonumber \\
& \quad   =- \int \prod_{n=1}^{N-1} d \bm{ \rho}'_{(ij), n}   e^{ -i \sum_{k=2}^{N-1}  \mathbf{ q}'_{(ij), k} \cdot    \bm{ \rho}'_{( ij),k}  }\nonumber \\
 & \quad \quad \quad \times       m V_{0}\delta(\bm{ \rho}'_{(ij), 1})    \psi_{[J]} ( \{ \bm{ \rho}'_{(ij)}  \} ,  \{ q_{(12)}  \})   ,
\end{align}
where   $\overline{\{  \mathbf{ q}'_{(ij)}  \}} = \{  \mathbf{ q}'_{(ij),2} ,  \mathbf{ q}'_{(ij),3} ,  \cdots ,  \mathbf{ q}'_{(ij),N-1}  \}$.
Using Eq.~(\ref{freewav2b}) and  Eq.~(\ref{renorm}), we find that 
 the wave function $\psi^{(ij)}_{[J]}$  and  $T^{(ij)}_{[J]} $ are related  by
 \begin{align}
 & \psi^{(ij)}_{[J]}( \{ \bm{ \rho}_{(ij)}  \} ,  \{q_{(12)}  \}) \nonumber \\
 & =    \int \prod_{n=2}^{N-1} \frac{ d \mathbf{ q}'_{(ij), n} }{(2\pi)^{3}}   e^{ i \sum_{k=2}^{N-1}  \mathbf{ q}'_{(ij), k} \cdot    \bm{ \rho}_{(ij), k}  }  \nonumber \\
& \times i h_{0}^{(+)} \left ( \sqrt{  \sigma^{2} - \sum_{n=2}^{N-1} \mathbf{ q}^{'2}_{(ij), n} }  \rho_{(ij),1}  \right )     \nonumber \\
 & \times  \frac{ \sqrt{  \sigma^{2} - \sum_{n=2}^{N-1} \mathbf{ q}^{'2}_{(ij), n} }}{4\pi}   T^{(ij)}_{[J]}   \left ( \overline{\{  \mathbf{ q}'_{(ij)}  \} },  \{  q_{(12)} \}  \right )  . \label{Twavfree}
\end{align}
The  $T^{(ij)}_{[J]} $ amplitudes are given by the solutions of coupled integral equations,
 \begin{align}
& T^{(ij)}_{[J]}   \left  (\overline{ \{  \mathbf{ q}'_{(ij)}  \} },  \{  q_{(12)} \}  \right  ) -   \mathbb{T}^{(ij)}_{[J]}   \left (\overline{ \{  \mathbf{ q}'_{(ij)}  \}} ,  \{  q_{(12)} \}  \right )\nonumber \\
&=    \frac{4\pi i t_{0} ( \sqrt{  \sigma^{2} - \sum_{n=2}^{N-1} \mathbf{ q}^{'2}_{(ij), n} } ) }{ \sqrt{  \sigma^{2} - \sum_{n=2}^{N-1} \mathbf{ q}^{'2}_{(ij), n} } }  \nonumber \\
 & \times i  \int \prod_{n=1}^{N-1} \frac{ d \mathbf{ q}''_{(ij), n} }{(2\pi)^{3}}  \frac{  \prod_{k=2}^{N-1} (2\pi)^{3}  \delta( \mathbf{ q}'_{( ij), k} - \mathbf{ q}''_{(ij), k} )  }{  \sigma^{2} - \sum_{n=1}^{N-1} \mathbf{ q}^{''2}_{(ij), n}  + i \epsilon}    \nonumber \\
 & \times   \sum^{N; (i'j'\neq ij)}_{(i'   < j' ) =1 } T^{(i'j')}_{[J]}   \left  (\overline{ \{  \mathbf{ q}''_{(i'j')}  \} },  \{  q_{(12)} \} \right   )  ,
 \end{align}
where
 \begin{align}
&   \mathbb{T}^{(ij)}_{[J]}  \left  (\overline{ \{  \mathbf{ q}'_{(ij)}  \}} ,  \{  q_{(12)} \} \right   ) \nonumber \\
&= \frac{ 4\pi t_{0}  ( \sqrt{  \sigma^{2} - \sum_{n=2}^{N-1} \mathbf{ q}^{'2}_{(ij), n} } ) }{ \sqrt{  \sigma^{2} - \sum_{n=2}^{N-1} \mathbf{ q}^{'2}_{(ij), n} }   }  \int \prod_{n=1}^{N-1} d \bm{ \rho}_{(ij), n}    \delta(\bm{ \rho}_{(ij),1})    \nonumber \\
 & \times  e^{ -i \sum_{k=2}^{N-1}  \mathbf{ q}'_{(ij), k} \cdot    \bm{ \rho}_{(ij), k}  }     \psi^{(0)}_{[J]}( \{ \bm{ \rho}_{(ij}  \} ,  \{ q_{(12)}  \})  .
\end{align}
The total $N$-body scattering amplitude is given by \mbox{$T_{[J]} = \sum_{(i<j)=1}^{N} T^{(ij)}_{[J]}$}.

\subsubsection{Construction of finite volume $N$-body wave functions } 
Given the solutions of the infinite-volume wave function, $\psi^{(ij)}$, using Eq.(\ref{fvrefwavconstr}),  the finite-volume wave function $\phi^{(ij)} $ is   obtained,
 \begin{align}  
& \phi_{[J]}^{(ij)}    ( \{ \bm{ \rho}_{(ij) }   \},  \{ q_{(12)} \})   \nonumber \\
& =    \sum_{ \{ \mathbf{ n}_{(ij)}  \}  \in \mathbb{Z}^{3} }   e^{i \frac{\mathbf{ P}}{N} \cdot    \left ( \sum_{k=1}^{N-1 } k \mathbf{ n}_{(ij),k}     \right ) L  }  \int \prod_{n=2}^{N-1} \frac{ d \mathbf{ q}'_{(ij), n} }{(2\pi)^{3}}    \nonumber \\
&     \times   e^{ i \sum_{n'=2}^{N-1}  \mathbf{ q}'_{(ij), n'} \cdot   \left ( \bm{ \rho}_{(ij), n'} + \sqrt{\frac{2}{n' (n'+1)}}  \sum_{k=1}^{n'}k \mathbf{ n}_{(ij),k} L \right )  }  \nonumber \\
& \times i h_{0}^{(+)} \left ( \sqrt{  \sigma^{2} - \sum_{n=2}^{N-1} \mathbf{ q}^{'2}_{(ij), n} }  |  \bm{ \rho}_{(ij),1} +\mathbf{ n}_{(ij),1}   L  |  \right )     \nonumber \\
 & \times   \frac{ \sqrt{  \sigma^{2} - \sum_{n=2}^{N-1} \mathbf{ q}^{'2}_{(ij), n} } }{4\pi}     T^{(ij)}_{[J]}   \left  (\overline{ \{  \mathbf{ q}'_{(ij)}  \} },  \{  q_{(12)} \}  \right  )  .
\end{align} 
Using the fact that
\begin{align}
& \sum_{n'=2}^{N-1}  \mathbf{ q}'_{(ij), n'} \cdot     \left (   \frac{ \sum_{k=2}^{n'} k \mathbf{ n'}_{(ij),k}   }{\sqrt{n'(n'+1)}}     \right ) L \nonumber \\
&  =  \sum_{k=2}^{N-1}  \left ( \sum_{n'=k}^{N-1}    \frac{ \mathbf{ q}'_{(ij), n'}   }{\sqrt{n'(n'+1)}}  \right ) \cdot     \left (      k \mathbf{ n}_{(ij),k}  \right ) L  ,
\end{align}
 and the Poisson summation formula, we find
 \begin{align}  
& \phi^{(ij)}_{[J]}    ( \{ \bm{ \rho }_{(ij) }   \},  \{ q_{(12)} \})   \nonumber \\
& =    \sum_{ \overline{\{ \mathbf{ n}_{(ij)} \}}  \in \mathbb{Z}^{3} }       \int \prod_{n=2}^{N-1} \frac{ d \mathbf{ q}'_{(ij), n} }{L^{3}}   e^{ i \sum_{n'=2}^{N-1}  \mathbf{ q}'_{(ij), n'} \cdot     \bm{ \rho}_{(ij), n'}   }  \nonumber \\
&     \times   \sum_{k=2}^{N-1} \delta      \left (\frac{k \mathbf{ P}}{N} + \sum_{n'=k}^{N-1} k  \sqrt{ \frac{  2 }{n'(n'+1)}} \mathbf{ q}'_{(ij), n'}  - \frac{2\pi}{L} \mathbf{ n}_{(ij),k}    \right )   \nonumber \\
& \times \sum_{\mathbf{ n}_{(ij),1} \in \mathbb{Z}^{3}}  e^{i   \left (    \frac{\mathbf{ P}}{N} +\sum_{n'=2}^{N-1}     \sqrt{\frac{2}{n' (n'+1)}}   \mathbf{ q}'_{(ij), n'}   \right )  \cdot    \mathbf{ n}_{(ij),1}    L  }  \nonumber \\
& \times i h_{0}^{(+)} \left ( \sqrt{  \sigma^{2} - \sum_{n=2}^{N-1} \mathbf{ q}^{'2}_{(ij), n} }  | \bm{ \rho}_{(ij),1} +\mathbf{ n}_{(ij),1}   L  |  \right )     \nonumber \\
 & \times   \frac{\sqrt{  \sigma^{2} - \sum_{n=2}^{N-1} \mathbf{ q}^{'2}_{(ij), n} } }{4\pi}      T^{(ij)}_{[J]}   \left  (\overline{ \{  \mathbf{ q}'_{(ij)}  \} },  \{  q_{(12)} \}  \right  )   \ . \label{fvphiijpwa}
\end{align}
 The total finite-volume wave function is given by \mbox{$ \phi_{[J]}    = \sum_{(i<j)=1}^{N} \phi^{(ij)}_{[J]}   $}.

Similarly, using Eq.~(\ref{overlinephiijpwa}) and Eq.~(\ref{Twavfree}), we also find
 \begin{align}
&   \chi^{(ij)}_{[J]}  ( \{  \bm{ \rho}_{(12)}  \} ,  \{ q_{(12)}  \})     = \chi^{ (0), (ij)}_{[J]}  ( \{ \bm{ \rho}_{(12)}  \} ,  \{ q_{(12)}  \})   \nonumber \\
& \quad \quad \quad \quad +   \sum^{N}_{(i'  < j')=1} \chi^{ (i' j'), (ij)}_{[J]}  ( \{  \bm{ \rho}_{(i'j')}  \} ,  \{ q_{(12)}  \})  , \label{Nbphiij}
\end{align}
where
 \begin{align}
&   \chi^{ (0), (ij)}_{[J]}  ( \{ \bm{ \rho}_{(12)}  \} ,  \{ q_{(12)}  \})    =     \sum_{ \overline{\{\mathbf{ n}_{(ij)} \}}  \in \mathbb{Z}^{3} }     e^{i \frac{\mathbf{ P}}{N} \cdot    \left ( \sum_{k=2}^{N-1 } k \mathbf{ n}_{ (ij),k}       \right ) L  }    \nonumber \\
&\quad \quad \quad \quad  \times  \psi^{(0)}_{[J]}   (\Gamma^{(12),(ij)}  \{ \overline{\ddot{ \bm{ \rho}}_{(ij)}}  \}   , \{ q_{(12)} \})   ,
\end{align}

 \begin{align}
  & \chi^{ (i j), (ij)}_{[J]}  ( \{  \bm{ \rho}_{(ij)}  \} ,  \{ q_{(12)}  \})    \nonumber \\
  &=  \sum_{ \overline{\{ \mathbf{ n}_{(ij)} \}}  \in \mathbb{Z}^{3} }       \int \prod_{n=2}^{N-1} \frac{ d \mathbf{ q}'_{(ij), n} }{L^{3}}   e^{ i \sum_{n'=2}^{N-1}  \mathbf{ q}'_{(ij), n'} \cdot     \bm{ \rho}_{(ij), n'}   }  \nonumber \\
&     \times   \sum_{k=2}^{N-1} \delta      \left (\frac{k \mathbf{ P}}{N} + \sum_{n'=k}^{N-1} k  \sqrt{ \frac{  2 }{n'(n'+1)}} \mathbf{ q}'_{(ij), n'}  - \frac{2\pi}{L} \mathbf{ n}_{(ij),k}    \right )   \nonumber \\
& \times i h_{0}^{(+)} \left ( \sqrt{  \sigma^{2} - \sum_{n=2}^{N-1} \mathbf{ q}^{'2}_{(ij), n} }   \rho_{(ij),1 }   \right )     \nonumber \\
 & \times    \frac{ \sqrt{  \sigma^{2} - \sum_{n=2}^{N-1} \mathbf{ q}^{'2}_{(ij), n} }}{4\pi}      T^{(ij)}_{[J]}   \left  (\overline{ \{  \mathbf{ q}'_{(ij)}  \} },  \{  q_{(12)} \}  \right  )    ,\label{Nbphiijij}
\end{align}
and
\begin{widetext}
 \begin{align}
   \chi^{ (i' j'), (ij)}_{[J]}  (  \{  \bm{ \rho}_{(i'j')}  \} , & \{ q_{(12)}  \})    \stackrel{(i' j') \neq (ij)}{ = }    \int \prod_{n=2}^{N-1} \frac{ d \mathbf{ q}'_{(i'j'), n} }{(2\pi)^{3}}    e^{ i \sum_{k=2}^{N-1}  \mathbf{ q}'_{(i'j'), k} \cdot      \bm{ \rho}_{(i'j'),k}     }  \nonumber \\
&  \quad \quad  \quad \quad  \times    \sum_{ \overline{\{\mathbf{ n}_{(ij)} \}}  \in \mathbb{Z}^{3} }      e^{ i   \sum_{k=2}^{N-1}  \left ( \frac{\mathbf{ P}}{N} +  \sum_{n=k}^{N-1}    \sqrt{ \frac{ 2  }{n(n+1)}}   \left (   \sum_{n'=2}^{N-1}   \mathbf{ q}'_{(i'j'), n'}  \Gamma^{(i' j'),(ij)}_{n', n}  \right )  \right ) \cdot     \left (      k \mathbf{ n}_{(ij),k}  \right ) L     }   \nonumber \\
& \quad \quad  \quad \quad  \times  i h_{0}^{(+)} \left ( \sqrt{  \sigma^{2} - \sum_{n=2}^{N-1} \mathbf{ q}^{'2}_{(i'j'), n} } \left |     \bm{ \rho}_{(i'j'),1}  +  \sum_{n=2}^{N-1}  \Gamma^{(i' j'),(ij)}_{1, n}  \sqrt{\frac{2}{n(n+1)}}  \sum_{k=2}^{n} k  \mathbf{ n}_{ (ij),k   } L  \right |   \right )    \nonumber \\
& \quad \quad  \quad \quad   \times\frac{ \sqrt{  \sigma^{2} - \sum_{n=2}^{N-1} \mathbf{ q}^{'2}_{(i'j'), n} }}{4\pi}   T^{(i'j')}_{[J]}   \left ( \overline{\{  \mathbf{ q}'_{(i'j')}  \} },  \{  q_{(12)} \}  \right )     . \label{Nbphiijipjp}
\end{align}
\end{widetext}

\subsubsection{$N$-body  secular equations} 
Given all the ingredients of the finite-volume wave functions from Eq.~(\ref{fvphiijpwa}) up to Eq.~(\ref{Nbphiijipjp}), the discrete energy spectra are thus given by $N(N-1)/2$   secular equations,
\begin{align}
& \det \bigg [ \int \prod_{n=1}^{N-1} d \bm{\rho}_{(ij),n}  \phi^{*}_{[J']}    ( \{ \bm{ \rho }_{(12) }   \},  \{ q_{(12)} \})  m V_{0} \delta( \bm{\rho}_{(ij),1} ) \nonumber \\
& \times  \left [  \chi^{(ij)}_{[J]}  ( \{  \bm{ \rho}_{(12)}  \} ,  \{ q_{(12)}  \}) - \phi_{[J]}    ( \{ \bm{ \rho }_{(12) }   \},  \{ q_{(12)} \})  \right ] \bigg ] =0. \label{nbodysecular}
\end{align}
As \mbox{$ \bm{\rho}_{(ij),1}  \rightarrow \mathbf{ 0}$},  both $\chi^{ (i j), (ij)}_{[J]} $ and $\phi^{(ij)}_{[J]}  $ appear to have the exact same ultravioletly divergent behavior because of the spherical Hankel function, $h_{0}^{(+)} ( \sqrt{  \sigma^{2} - \sum_{n=2}^{N-1} \mathbf{ q}^{'2}_{(ij), n} }   \rho_{(ij),1 }   ) \sim \frac{1}{\rho_{(ij),1}} $, thus, the subtraction of two terms is completely free from ultraviolet divergence. The ultraviolet divergence   appears in $ \phi^{*}_{[J']}  $ as well. Since $V_{0}$ is a bare parameter, in order to remove all the ultraviolet divergence, a renormalization of the energy is required. Notice the fact that
\begin{align}
mV_{0} i h_{0}^{(+)} (q r) \stackrel{ r \rightarrow 0 }{  \rightarrow} \frac{4\pi}{q} -  \frac{ m V_{0} }{t_{0}(q)},
\end{align}
after the shift on the energy: $E \rightarrow E + \delta E$, the divergent contribution of secular equation can be completely canceled out by the counter term, $\delta E$. Thus the renormalized quantity $ \phi^{*}_{[J']}   m V_{0}$ may be given by
 \begin{align}  
&   \phi^{*}_{[J']}    ( \{ \bm{ \rho }_{(12) }   \},  \{ q_{(12)} \})  m V_{0}    \nonumber \\
& \stackrel{ \bm{\rho}_{(ij),1} \rightarrow \mathbf{ 0}}{ \rightarrow   }   \sum_{ \overline{\{ \mathbf{ n}_{(ij)} \}}  \in \mathbb{Z}^{3} }       \int \prod_{n=2}^{N-1} \frac{ d \mathbf{ q}'_{(ij), n} }{L^{3}}   e^{ -i \sum_{n'=2}^{N-1}  \mathbf{ q}'_{(ij), n'} \cdot     \bm{ \rho}_{(ij), n'}   }  \nonumber \\
&     \times   \sum_{k=2}^{N-1} \delta      \left (\frac{k \mathbf{ P}}{N} + \sum_{n'=k}^{N-1} k  \sqrt{ \frac{  2 }{n'(n'+1)}} \mathbf{ q}'_{(ij), n'}  - \frac{2\pi}{L} \mathbf{ n}_{(ij),k}    \right )   \nonumber \\
& \times     T^{(ij)* }_{[J']}   \left  (\overline{ \{  \mathbf{ q}'_{(ij)}  \} },  \{  q_{(12)} \}  \right  )   . \label{renormleftseq}
\end{align}

\subsection{Three-body interaction with $\delta$-function potential}

The three-body solutions can be derived from $N$-body results given in Section \ref{Nbwav} by setting \mbox{$N=3$}. Since the number of particles in the three-body  problem is still manageable,  we use the short-hand notation,
\begin{align}
&\mathbf{ r}_{(ij)} = \bm{\rho}_{(ij),1}, \ \ \ \ \mathbf{ r}_{k} = \bm{\rho}_{(ij),2} , \nonumber \\
& \mathbf{ q}_{(ij)} = \mathbf{ q}_{(ij),1},  \ \ \ \  \mathbf{ q}_{k} = \mathbf{ q}_{(ij),2},  \ \ i\neq j \neq k.
\end{align}
The different sets are related by linear transformations, such as
\begin{align}
& \mathbf{ r}_{(13)} = \frac{\mathbf{ r}_{(12)} + \sqrt{3} \mathbf{ r}_{3}}{2}  , \  \mathbf{ r}_{2} = \frac{\sqrt{3}  \mathbf{ r}_{(12)}-  \mathbf{ r}_{3} }{2} ,  \nonumber \\
& \mathbf{ q}_{(13)} = \frac{\mathbf{ q}_{(12)} + \sqrt{3} \mathbf{ q}_{3}}{2}  , \   \mathbf{ q}_{2} = \frac{\sqrt{3}  \mathbf{ q}_{(12)}-  \mathbf{ q}_{3} }{2} .
\end{align}

 According to Eq.~(\ref{Twavfree}), the  infinite-volume wave function, $ \psi^{(ij)}_{[J]}$, is given by
 \begin{align}
 & \psi^{(ij)}_{[J]}( \mathbf{ r}_{(ij)} , \mathbf{ r}_{k}  ; q_{(12)} , q_{3})   \nonumber \\
 & =    \int  \frac{ d \mathbf{ q}  }{(2\pi)^{3}}     i h_{0}^{(+)} \left ( \sqrt{  \sigma^{2} -   \mathbf{ q}^{2}  } r_{(ij)}  \right )   e^{ i   \mathbf{ q}  \cdot    \mathbf{ r}_{k}  }     \nonumber \\
 & \quad  \times     \frac{  \sqrt{  \sigma^{2} -  \mathbf{ q}^{2} } }{4\pi}    T^{(ij)}_{[J]} \left  (  \mathbf{ q}   ; q_{(12)} , q_{3} \right  )  ,  
\end{align}
where the solutions of the $T$-amplitude are given by
\begin{align}
& T^{(ij)} \left  (  \mathbf{ q}  ; \mathbf{ q}_{(12)} , \mathbf{ q}_{3} \right  ) - \mathbb{T}^{(ij)} \left (  \mathbf{ q}  ; \mathbf{ q}_{(12)} , \mathbf{ q}_{3} \right )\nonumber \\
&=    \frac{4\pi i t_{0} ( \sqrt{  \sigma^{2} -  \mathbf{ q}^{2} } ) }{ \sqrt{  \sigma^{2} -  \mathbf{ q}^{2}  } }  i  \int  \frac{ d \mathbf{ q}'  }{(2\pi)^{3}}  \nonumber \\
 &  \quad  \times     \frac{     T_{(ik  )} \left  ( \frac{\sqrt{3}}{2}  \mathbf{ q'}   ; \mathbf{ q}_{(12)} , \mathbf{ q}_{3}  \right   ) + T_{(  jk  )} \left  ( \frac{\sqrt{3}}{2}  \mathbf{ q'}   ;\mathbf{ q}_{(12)} , \mathbf{ q}_{3} \right   ) }{  \sigma^{2}-  \frac{3}{4} \mathbf{ q}^{' 2}  -( \frac{ \mathbf{ q'} }{2}+  \frac{2}{\sqrt{3}} \mathbf{ q}  )^{2}    + i \epsilon}    ,  
 \end{align}
 where
  \begin{align}
   \mathbb{T}^{(ij)}_{[J]} & \left  (   \mathbf{ q}   ;   q_{(12)} , q_{3}  \right   )  
 = \frac{ 4\pi t_{0} ( \sqrt{  \sigma^{2} -   \mathbf{ q}^{2} } ) }{ \sqrt{  \sigma^{2} -  \mathbf{ q}^{2}  }   }      \nonumber \\
 & \times   \int  d \mathbf{ r}_{k}   e^{ -i    \mathbf{ q}  \cdot    \mathbf{ r}_{k}  }     \psi^{(0)}_{[J]}(  \mathbf{ 0},  \mathbf{ r}_{k}  ;  q_{(12)} , q_{3} )  .
\end{align}
 
In terms of the $T$-amplitude, the finite volume three-body wave function is taken from Eq.~(\ref{fvphiijpwa}),
 \begin{align}  
& \phi^{(ij)}_{[J]}     ( \mathbf{ r}_{(ij) } ,  \mathbf{ r}_{k }  ; q_{(12)} , q_{3})  =   \frac{(\frac{\sqrt{3}}{2})^{3}}{L^{3}} \sum_{  \mathbf{ n}   \in \mathbb{Z}^{3} }^{  \mathbf{ q}  =- \frac{\mathbf{ P}}{\sqrt{3}} + \frac{\sqrt{3}}{2}  \frac{2\pi}{L} \mathbf{ n}   }        e^{ i   \mathbf{ q} \cdot     \mathbf{ r}_{k}   }  \nonumber \\
& \times \sum_{\mathbf{ n}_{(ij)}  \in \mathbb{Z}^{3}  } e^{i   \left (    \frac{\mathbf{ P}}{3} +      \frac{ \mathbf{ q}  }{\sqrt{3}}    \right )  \cdot    \mathbf{ n}_{(ij)}    L  }   i  h_{0}^{(+)} \left ( \sqrt{  \sigma^{2} -  \mathbf{ q}^{2}  }  |  \mathbf{ r}_{(ij)} +\mathbf{ n}_{(ij)}   L  |  \right )     \nonumber \\
 & \times     \frac{ \sqrt{  \sigma^{2} -   \mathbf{ q}^{2} } }{4\pi}    T^{(ij)}_{[J]}   \left  (  \mathbf{ q}; q_{(12)} , q_{3})  \right  )  . \label{fv3bphiijpwa}
\end{align}

Similarly, from Eq.~(\ref{Nbphiij}), Eq.~(\ref{Nbphiijij}), and Eq.~(\ref{Nbphiijipjp}),  we also find
 \begin{align}
&   \chi^{(ij)}_{[J]}  ( \mathbf{ r}_{(12) } ,  \mathbf{ r}_{3 }  ;   q_{(12)} , q_{3} )        \nonumber \\
& \quad \quad \quad   =   \sum^{N}_{(i'  < j')=1}  \chi^{ (i' j'), (ij)}_{[J]}  (  \mathbf{ r}_{(i'j') } ,  \mathbf{ r}_{k' } ; q_{(12)} , q_{3} )  ,
\end{align}
where
 \begin{align}
&  \chi^{ (i j), (ij)}_{[J]}  (  \mathbf{ r}_{(ij) } ,  \mathbf{ r}_{k } ; q_{(12)} , q_{3}  )         =     \frac{(\frac{\sqrt{3}}{2})^{3}}{L^{3}} \sum_{  \mathbf{ n}   \in \mathbb{Z}^{3} }^{  \mathbf{ q}  =- \frac{\mathbf{ P}}{\sqrt{3}} + \frac{\sqrt{3}}{2}  \frac{2\pi}{L} \mathbf{ n}   }        e^{ i   \mathbf{ q} \cdot     \mathbf{ r}_{k}   }     \nonumber \\
& \times  i   h_{0}^{(+)} \left ( \sqrt{  \sigma^{2} -   \mathbf{ q}^{2}  } r_{(ij)}   \right )     \frac{ \sqrt{  \sigma^{2} -   \mathbf{ q}^{2}  }}{4\pi}   T^{(ij)}_{[J]} \left (    \mathbf{ q}   ; q_{(12)} , q_{3} \right )    ,
\end{align}
and
 \begin{align}
&   \chi^{ (i' j'), (ij)}_{[J]}  (  \mathbf{ r}_{(i'j') } ,  \mathbf{ r}_{k' } ; q_{(12)} , q_{3}  )         \nonumber \\
&   \stackrel{(i'j') \neq (ij) }{  = }      \int  \frac{ d \mathbf{ q}  }{(2\pi)^{3}}   e^{ i   \mathbf{ q}  \cdot    \mathbf{ r}_{k'}    }     \sum_{ \mathbf{ n}   \in \mathbb{Z}^{3} }        e^{ i  \left ( \frac{2\mathbf{ P}}{3} +   \frac{2\Gamma^{(i' j'),(ij)}_{2,2}   }{\sqrt{3}}  \mathbf{ q}  \right )  \cdot            \mathbf{ n}      L      }     \nonumber \\
& \quad \quad  \times  i   h_{0}^{(+)} \left ( \sqrt{  \sigma^{2} -   \mathbf{ q}^{2}  }  \left | \mathbf{ r}_{(i'j')} +   \frac{   2  \Gamma^{(i' j'),(ij)}_{1,2}  }{\sqrt{3}}   \mathbf{ n}     L  \right |  \right )  \nonumber \\
& \quad \quad \times     \frac{ \sqrt{  \sigma^{2} -   \mathbf{ q}^{2}  }}{4\pi}   T^{(i'j')}_{[J]} \left (    \mathbf{ q}   ; q_{(12)} , q_{3} \right )    .
\end{align}

For the three-body problem, three secular equations may be obtained according to Eq.~(\ref{nbodysecular}),  but only two of them are independent. For example, for channel $(12)$,  the quantization condition is given by
\begin{align}
 \det \bigg [ &  \int  d \mathbf{ r}_{3}   \phi^{*}_{[J']}     ( \mathbf{ 0} ,  \mathbf{ r}_{3 }  ; q_{(12)} , q_{3})  m V_{0} \nonumber \\
& \times  \triangle   \phi^{(12)}_{[J]}     ( \mathbf{ 0} ,  \mathbf{ r}_{3 }  ; q_{(12)} , q_{3})  \bigg ] =0,
\end{align}
where  
 \begin{align} 
&   \triangle   \phi^{(12)}_{[J]}      ( \mathbf{ 0} ,  \mathbf{ r}_{3 }  ; q_{(12)} , q_{3})  \nonumber \\
& =       -      \frac{(\frac{\sqrt{3}}{2})^{3}}{L^{3}} \sum_{  \mathbf{ n}_{3}   \in \mathbb{Z}^{3} }^{  \mathbf{ q}  =- \frac{\mathbf{ P}}{\sqrt{3}} + \frac{\sqrt{3}}{2}  \frac{2\pi}{L} \mathbf{ n}_{3}   }   \frac{ \sqrt{  \sigma^{2} -   \mathbf{ q}^{2}  }}{4\pi}      T^{(12)}_{[J]}   \left  (  \mathbf{ q}; q_{(12)} , q_{3}  \right  )      \nonumber \\
&\quad \quad  \times  \left [ \sum_{\mathbf{ n}   \in \mathbb{Z}^{3}  }^{ \mathbf{ n} \neq \mathbf{ 0} } e^{i   \left (    \frac{\mathbf{ P}}{3} +      \frac{ \mathbf{ q}  }{\sqrt{3}}    \right )  \cdot    \mathbf{ n}     L  }   i  h_{0}^{(+)} \left ( \sqrt{  \sigma^{2} -  \mathbf{ q}^{2}  }  | \mathbf{ n}  L  |  \right )   \right ]    e^{ i   \mathbf{ q} \cdot     \mathbf{ r}_{3}   }    \nonumber \\ 
& +     \bigg [  \int  \frac{ d \mathbf{ q}  }{(2\pi)^{3}}      \sum_{ \mathbf{ n}   \in \mathbb{Z}^{3} }      e^{  i  \left ( \frac{2\mathbf{ P}}{3} -   \frac{ \mathbf{ q}  }{\sqrt{3}}  \right )  \cdot            \mathbf{ n}      L      }    \nonumber \\
 & \quad \quad   -   \frac{(\frac{\sqrt{3}}{2})^{3}}{L^{3}} \sum_{  \mathbf{ n}_{k}   \in \mathbb{Z}^{3} }^{  \mathbf{ q}  =- \frac{\mathbf{ P}}{\sqrt{3}} + \frac{\sqrt{3}}{2}  \frac{2\pi}{L} \mathbf{ n}_{k}   }            \sum_{\mathbf{ n}  \in \mathbb{Z}^{3}  } e^{i   \left (    \frac{\mathbf{ P}}{3} +      \frac{ \mathbf{ q}  }{\sqrt{3}}    \right )  \cdot    \mathbf{ n}  L  }   \bigg ]  \nonumber \\
 & \quad \quad  \times  i   h_{0}^{(+)} \left ( \sqrt{  \sigma^{2} -  \mathbf{ q}^{2}  }  |  \frac{ \sqrt{3} \mathbf{ r}_{3} }{2} +\mathbf{ n}  L  |  \right )  e^{  - i   \mathbf{ q}  \cdot   \frac{  \mathbf{ r}_{3}  }{2}  } \nonumber \\
 &\quad \quad  \times     \frac{ \sqrt{  \sigma^{2} -   \mathbf{ q}^{2}  }}{4\pi}   \sum_{(ij)  = 13}^{23}   T^{(ij)}_{[J]}   \left  (  \mathbf{ q}; q_{(12)} , q_{3} \right  )  .
\end{align}
The renormalized $ \phi^{*}_{[J']}  m V_{0}$ may be given by
 \begin{align}  
&   \phi^{*}_{[J]}    ( \mathbf{ 0} ,  \mathbf{ r}_{3 }  ; q_{(12)} , q_{3} )  m V_{0}    \nonumber \\
& \rightarrow      \frac{(\frac{\sqrt{3}}{2})^{3}}{L^{3}} \sum_{  \mathbf{ n}   \in \mathbb{Z}^{3} }^{  \mathbf{ q}  =- \frac{\mathbf{ P}}{\sqrt{3}} + \frac{\sqrt{3}}{2}  \frac{2\pi}{L} \mathbf{ n}   }        e^{ - i   \mathbf{ q} \cdot     \mathbf{ r}_{3}   }       T^{(12)* }_{[J']}   \left  (  \mathbf{ q}  ;  q_{(12)},q_{3}  \right  )   .  
\end{align}

Normally, the infinite sum of Bessel functions in the above equation has poor convergence, and for   numerical  purposes a better expression may be given by Eq.~(\ref{luscherlatsum}) in a partial-wave expanded form. It may be also convenient to use an alternative form in Eq.(\ref{latsumfast}) without partial wave expansion, 
 \begin{equation}
  \sum_{ \mathbf{ n }  \in \mathbb{Z}^{3} }  e^{i   \mathbf{ Q}    \cdot   \mathbf{ n} L   }      h_{0}^{(+)} (q    | \mathbf{ r} +  \mathbf{ n}  L   |)      =  \mathcal{J}^{(  \mathbf{ Q}   )}  ( \mathbf{ r}, q)  + i n_{0} (q r) ,   \nonumber \\
\end{equation}
where the expression and derivation of $\mathcal{J}^{(  \mathbf{ Q}   )} ( \mathbf{ r}, q)  $ is presented  in Section \ref{latsum2bgreen}.  In terms of a partial-wave expanded form, $\mathcal{J}^{(  \mathbf{ Q}   )}    $  is related to L\"uscher's form by
\begin{equation}
 i  \mathcal{J}^{(  \mathbf{ Q}   )}  ( \mathbf{ r}, q)   =  \sqrt{4\pi} \sum_{[j]}  \mathcal{M}_{[0] ,[j]}^{(\mathbf{ Q})}  (q) j_{j} (q r ) Y_{[j]} (\mathbf{ r})   .
\end{equation}

\section{Lattice sum of two-body Green's function} \label{latsum2bgreen}
An alternative fast algorithm of  performing the lattice sum of the two-body Green's function in Eq.~(\ref{fvphiijpwa}) and Eq.~(\ref{fv3bphiijpwa}) without partial-wave expansion is provided in this section.   First of all, using the identity
\begin{align}
\frac{q}{4\pi} i h_{0}^{(+)} (q r ) = \frac{1}{2\pi \sqrt{\pi}} \int_{0}^{\infty} d t e^{ - r^{2} t^{2} + \frac{q^{2}}{4 t^{2}}}, 
\end{align}
and also splitting   the integration by an arbitrary parameter, $\eta$, we can rewrite the lattice sum of the Green's function in Eq.~(\ref{luscherlatsum})   as
 \begin{align}
&  \sum_{ \mathbf{ n }  \in \mathbb{Z}^{3} }  e^{i    \mathbf{ Q}   \cdot   \mathbf{ n} L   }      h_{0}^{(+)} (q    | \mathbf{ r} +   \mathbf{ n}  L   |)     \nonumber \\
& = - \frac{2 i}{\sqrt{\pi} q}  \sum_{ \mathbf{ n }  \in \mathbb{Z}^{3} }  e^{i   \mathbf{ Q}    \cdot   \mathbf{ n} L   }  \left [    \int_{0}^{\eta}  + \int_{\eta}^{\infty} \right ] d t e^{ -   | \mathbf{ r} + \mathbf{ n}  L   |^{2} t^{2} + \frac{q^{2}}{4 t^{2}}}. \label{latsum}
\end{align}
For the first term in Eq.(\ref{latsum}), using the identity
\begin{equation}
\frac{1}{2\pi \sqrt{\pi}} e^{- r^{2} t^{2}} =\frac{1}{2 t^{3}}   \int \frac{d \mathbf{ q'}}{(2\pi)^{3}} e^{ - \frac{ \mathbf{ q'}^{2} }{4 t^{2}}} e^{i \mathbf{ q'} \cdot \mathbf{ r}},
\end{equation}
and also applying Poisson summation, we find
 \begin{align}
&    - \frac{2 i}{\sqrt{\pi} q}  \sum_{ \mathbf{ n }  \in \mathbb{Z}^{3} }  e^{i   \mathbf{ Q}    \cdot   \mathbf{ n} L   }     \int_{0}^{\eta} d t e^{ -   | \mathbf{ r} +   \mathbf{ n}  L   |^{2} t^{2} + \frac{q^{2}}{4 t^{2}}} \nonumber \\
& = \frac{4\pi i}{q}  \frac{1}{L^{3}}   \sum_{ \mathbf{ n} \in \mathbb{Z}^{3}}^{\mathbf{ q'} =  \frac{2   \pi}{L} \mathbf{ n} -   \mathbf{ Q}   }  \frac{ e^{ \frac{ q^{2} - \mathbf{ q'}^{2}  }{4 \eta^{2}}}  }{q^{2} - \mathbf{ q'}^{2} + i \epsilon}   e^{i \mathbf{ q'}  \cdot \mathbf{ r}  }  .
\end{align}
For the second term in  Eq.~(\ref{latsum}), except for $\mathbf{ n}=\mathbf{ 0}$, the convergence of the integration is well-defined for a finite  value  of $\eta$, thus, we would like to isolate the  $\mathbf{ n}=\mathbf{ 0}$ piece, with the help of the identity
\begin{align}
& - \frac{2 i}{\sqrt{\pi} q}    \int_{\eta}^{\infty}  d t e^{ -  r^{2} t^{2} + \frac{q^{2}}{4 t^{2}}}  \nonumber \\
& =  i n_{0} (q r)  + i \frac{e^{- i q r} \mbox{erf} ( - \frac{i q}{2 \eta} + r \eta )  -  e^{i q r} \mbox{erf} ( - \frac{i q}{2 \eta}  -  r \eta )  }{2 q r} \ .
\end{align}
Therefore, we find
 \begin{align}
&  - \frac{2 i}{\sqrt{\pi} q}  \sum_{ \mathbf{ n }  \in \mathbb{Z}^{3} }  e^{i   \mathbf{ Q}    \cdot   \mathbf{ n} L   }   \int_{\eta}^{\infty}   d t e^{ -   | \mathbf{ r} +   \mathbf{ n}  L   |^{2} t^{2} + \frac{q^{2}}{4 t^{2}}} \nonumber \\
& =  i n_{0} (q r)  + i \frac{e^{- i q r} \mbox{erf} ( - \frac{i q}{2 \eta} + r \eta )  -  e^{i q r} \mbox{erf} ( - \frac{i q}{2 \eta}  -  r \eta )  }{2 q r} \nonumber \\
&  - \frac{2 i}{\sqrt{\pi} q}  \sum_{ \mathbf{ n }  \in \mathbb{Z}^{3} }^{ \mathbf{ n} \neq \mathbf{ 0}}  e^{i   \mathbf{ Q}   \cdot   \mathbf{ n} L   }   \int_{\eta}^{\infty}   d t e^{ -   | \mathbf{ r} +   \mathbf{ n}  L   |^{2} t^{2} + \frac{q^{2}}{4 t^{2}}}.
\end{align}
Putting everything together, we thus obtain  a fast convergent expression of the lattice sum of the Green's function without partial wave expansion,
 \begin{equation}
  \sum_{ \mathbf{ n }  \in \mathbb{Z}^{3} }  e^{i   \mathbf{ Q}    \cdot   \mathbf{ n} L   }      h_{0}^{(+)} (q    | \mathbf{ r} +  \mathbf{ n}  L   |)      =  \mathcal{J}^{(  \mathbf{ Q}   )}  ( \mathbf{ r}, q)  + i n_{0} (q r) , \label{latsumfast}
\end{equation}
where
 \begin{align}
 & \mathcal{J}^{(  \mathbf{ Q}   )}  ( \mathbf{ r}, q)   
 =   \frac{4\pi i}{q}  \frac{1}{L^{3}}   \sum_{ \mathbf{ n} \in \mathbb{Z}^{3}}^{\mathbf{ q'} =  \frac{2   \pi}{L} \mathbf{ n} -  \mathbf{ Q}   }  \frac{ e^{ \frac{ q^{2} - \mathbf{ q'}^{2}  }{4 \eta^{2}}}  }{q^{2} - \mathbf{ q'}^{2} + i \epsilon}   e^{i \mathbf{ q'}  \cdot \mathbf{ r}  }  \nonumber \\
&   + i \frac{e^{- i q r} \mbox{erf} ( - \frac{i q}{2 \eta} + r \eta )  -  e^{i q r} \mbox{erf} ( - \frac{i q}{2 \eta}  -  r \eta )  }{2 q r} \nonumber \\
&  - \frac{2 i}{\sqrt{\pi} q}  \sum_{ \mathbf{ n }  \in \mathbb{Z}^{3} }^{ \mathbf{ n} \neq \mathbf{ 0}}  e^{i   \mathbf{ Q}   \cdot   \mathbf{ n} L   }   \int_{\eta}^{\infty}   d t e^{ -   | \mathbf{ r} +  \mathbf{ n}  L   |^{2} t^{2} + \frac{q^{2}}{4 t^{2}}}.
\end{align}

\end{document}